\documentclass[10pt, letter, onecolumn]{arxiv}

\usepackage{kantlipsum, lipsum}
\usepackage{dm-colors}
\usepackage{amsmath}
\usepackage{pstricks, pst-node}
\usepackage{verbatim}
\usepackage{multirow}
\usepackage{scalerel}
\usepackage{booktabs}
\usepackage{enumitem}
\usepackage{xspace}
\usepackage{bm}
\usepackage{bbm}
\usepackage{mathtools}
\usepackage{soul}
\usepackage{epsfig}
\usepackage{graphicx}
\usepackage{amssymb}
\usepackage{colortbl}
\usepackage{csquotes}
\usepackage{setspace}
\usepackage{colortbl}
\usepackage{tabularx,ragged2e}
\usepackage{placeins}
\usepackage[symbol]{footmisc}
\usepackage{nameref}
\usepackage{varioref}
\usepackage[pagebackref=false,breaklinks=false,%
            colorlinks=true,bookmarks=true,citecolor=ourdarkblue,%
            urlcolor=ourdarkblue,linkcolor=ourdarkblue]{hyperref}
\usepackage[capitalize]{cleveref}
\usepackage{etoc}

\graphicspath{{figures/}}

\definecolor{purple}{RGB}{100,0,200}

\def\eg{{\em e.g.,}}
\def\ie{{\em i.e.,}}

\def\etc{{\em etc}\xspace}

\title{\Large{Exploring Recommendation Capabilities of GPT-4V(ision): \\ A Preliminary Case Study}}

\vspace{1cm}
\author[1$\ast$]{Peilin Zhou} 
\author[2$\ast$]{Meng Cao}
\author[1$\ast$]{You-Liang Huang}
\author[3$\ast$]{Qichen Ye}
\author[4]{Peiyan Zhang}
\author[5]{\\ \vspace{0.12cm}Junling Liu}
\author[4]{Yueqi Xie}
\author[6]{Yining Hua}
\author[4]{Jaeboum Kim}
\affil[1]{\normalsize Hong Kong University of Science and Technology (Guangzhou) \authorcr \hspace{8pt}}

\affil[2]{\normalsize International Digital Economy Academy (IDEA)  \vspace{3pt} 
}

\affil[3]{\normalsize Peking University \authorcr 
}
\affil[4]{\normalsize Hong Kong University of Science and Technology \authorcr 
}
\affil[5]{\normalsize Alibaba  \authorcr 
}
\affil[6]{\normalsize Harvard T.H. Chan School of Public Health  \authorcr }

\renewcommand{\correspondingauthor}[1]{$\ast$~Equal Contribution. \\
Email: \{zhoupalin\}@gmail.com, yeeeqichen@pku.edu.cn, yhuang142@connect.hkust-gz.edu.cn}

\begin{document}
%\begin{refsection}

\begin{abstract}
\section*{\centering Abstract}
Large Multimodal Models (LMMs) have demonstrated impressive performance across various vision and language tasks, yet their potential applications in recommendation tasks with visual assistance remain unexplored. 
To bridge this gap, we present a preliminary case study investigating the recommendation capabilities of GPT-4V(ison), a recently released LMM by OpenAI.
We construct a series of qualitative test samples spanning multiple domains and employ these samples to assess the quality of GPT-4V's responses within recommendation scenarios.
Evaluation results on these test samples prove that GPT-4V has remarkable zero-shot recommendation abilities across diverse domains, thanks to its robust visual-text comprehension capabilities and extensive general knowledge. However, we have also identified some limitations in using GPT-4V for recommendations, including a tendency to provide similar responses when given similar inputs.
This report concludes with an in-depth discussion of the challenges and research opportunities associated with utilizing GPT-4V in recommendation scenarios. Our objective is to explore the potential of extending LMMs from vision and language tasks to recommendation tasks. We hope to inspire further research into next-generation multimodal generative recommendation models, which can enhance user experiences by offering greater diversity and interactivity. All images and prompts used in this report will be accessible at \url{https://github.com/PALIN2018/Evaluate_GPT-4V_Rec}.
\end{abstract}

\maketitle

% ------------ SECTIONS ---------------------

\tableofcontents

\newpage

\listoffigures

\newpage

% ========== Edit your name here
\section{Introduction}
% peilin and yueqi
Large Language Models (LLMs) have recently witnessed remarkable advancements, revolutionizing natural language understanding and generation~\cite{zhao2023survey}. 
ChatGPT~\cite{openai2023gpt4}, a prominent and pioneer LLM, has attracted significant attention across both academia and industry.
Related research and applications span a wide spectrum, from text classification~\cite{loukas2023breaking, shi2023chatgraph, reiss2023testing}, sentiment analysis~\cite{sudirjo2023application, wang2023chatgpt,susnjak2023applying}, and named entity recognition~\cite{wei2023zero, hu2023zero} to content generation~\cite{benoit2023chatgpt, cao2023comprehensive} and chatbots~\cite{tlili2023if, shahsavar2023role, alshurafat2023usefulness, chow2023impact, panda2023exploring, ye2023qilin}, among others.

Given the textual reasoning capacity of ChatGPT, researchers have extensively examined its potential to enhance Recommender Systems (RS)~\cite{liu2023chatgpt, gao2023chat, lin2023sparks, nastasi2023does, di2023evaluating, zhang2023chatgpt, di2023retrieval, li2023preliminary}.
For example, the pioneering study by Liu et al. ~\cite{liu2023chatgpt} establishes a benchmark for evaluating ChatGPT’s performance in various recommendation tasks and compares it with traditional recommendation models. Their findings highlight ChatGPT's effectiveness in generating explanations and review summaries.
Di et al. ~\cite{di2023retrieval} performed a preliminary exploration of ChatGPT’s recommendation capabilities and discovered that ChatGPT could achieve comparable performance to state-of-the-art recommendation system models, even without the need for fine-tuning or prompt-engineering techniques.
Gao et al. ~\cite{gao2023chat} integrated ChatGPT with conventional recommender systems using prompts, demonstrating how this integration can enhance interactivity, comprehensibility, and cross-domain recommendation capabilities.
Additionally, Li et al. ~\cite{li2023preliminary} assessed ChatGPT's effectiveness in personalizing news recommendations and detecting fake news, observing that JSON format is more effective than textual representation when dealing with lengthy prompts.
These investigations have collectively demonstrated the effectiveness of ChatGPT in single-modal recommendation scenarios, primarily relying on textual information to make informed recommendations. However, the ever-evolving landscape of recommendation systems has increasingly demanded more sophisticated solutions capable of handling multi-modal data, encompassing not only textual information but also visual and other sensory inputs. In response, multimodal recommendation systems have gained prominence, leveraging information fusion from different modalities, such as text, images, audio, and more, to enhance the quality and relevance of recommendations~\cite{zhou2023comprehensive, liu2023multimodal}.

These systems can benefit from the rapid development of Large Multimodal Models (LMMs), which can encode multi-modal information within a unified semantic space~\cite{xu2023multimodal, liu2023qilin}. One such notable LMM is GPT-4V~\cite{gpt4v}, recently released by OpenAI. GPT-4V builds upon the state-of-the-art LLM GPT-4 \cite{openai2023gpt4} and is further trained on a comprehensive multimodal dataset, enhancing ChatGPT's capacity to understand visual data and extending its potential applications. While GPT-4V's performance has been examined in various visual-language tasks, including video understanding~\cite{lin2023mm}, optical character recognition (OCR)~\cite{shi2023exploring}, and image context reasoning~\cite{liu2023hallusionbench}, no research has yet explored its utility in recommendation scenarios. Thus, a fundamental question arises: \textbf{Can GPT-4V serve as an effective recommender system?} This question holds significant importance for the recommendation research community, especially within the emerging field of multimodal generative recommendations, providing valuable insights and guidance for future exploration.

\subsection{Motivation}
\label{sec:motivation}
% peilin and yueqi
Our main objective is to assess GPT-4V's performance as a recommender system. To achieve this, we have conducted a series of case studies across various recommendation domains such as culture, art, media, entertainment, and retail.

Through these case studies, we aim to evaluate GPT-4V's adaptability and potential in recommending items by leveraging both textual and visual information. These studies will provide insights into the strengths and limitations of GPT-4V as a multi-modal recommender system.

Our exploration of GPT-4V is guided by three types of research questions:
\begin{enumerate}
\item Input Modes
\begin{itemize}
\item \textbf{Image w/o description:} Can GPT-4V provide meaningful recommendations without being provided with a description of the image content and only given the image and task instructions?
\item \textbf{Image w/ description:} Will the recommendation results be more in line with user preferences when GPT-4V receives images along with their descriptions? Can GPT-4V accurately capture subtle details in images and combine them with text for accurate recommendations? How does GPT-4V weigh and decide when there is conflicting or inconsistent information between the image and text? 
% \item \textbf{Alternate text and image:} How does GPT-4V balance and integrate both textual and visual information when they appear alternately in a conversation for recommendations?
\item \textbf{Multi-turn dialogues:} In multi-turn dialogues, can the model effectively maintain the context for coherent recommendations?
\item \textbf{Multiple images:} When given multiple images, can GPT-4V integrate and analyze them to give a comprehensive recommendation? How does GPT-4V determine the priority of various images in multi-image inputs and make recommendations accordingly?
\end{itemize}
\item Input Complexity and Quality
\begin{itemize}
\item \textbf{Low-quality input:} When the quality of the input image is low (e.g., blurred), and when the text contains unclear grammar or spelling errors, can GPT-4V still accurately comprehend the content of the image and provide reasonable recommendations?
% \item \textbf{Cross-domain input:} Can GPT-4V make cross-domain recommendations?
% \item \textbf{Image dynamic:} Can GPT-4V, when given an image representing a dynamic scene, combine it with text to understand and recommend relevant products accurately?
% \item \textbf{Image style:} Can GPT-4V recognize different image styles and make recommendations accordingly?
\item  \textbf{Image sentiment:} Can GPT-4V capture sentiments in images and texts and make recommendations based on them?
\end{itemize}
\item Controllability, Interpretability, and Reliability
\begin{itemize}
\item  \textbf{Feedback-based adjustment:}  Can GPT-4V adjust its recommendation strategy after users provide feedback on its recommendations?
\item \textbf{Query refusion:} In what scenarios might GPT4-V refuse to make recommendations?
% \item \textbf{Hallucinations:} Under what circumstances would GPT4-V exhibit hallucinations in its recommendation results?
\item \textbf{Explainations:} Can GPT4-V provide reasonable explanations for its recommendations?
% \item \textbf{Bias:} Are there bias and fairness issues when using GPT4-V for recommendations?
% \item \textbf{Consistency:} When given similar image and text inputs, does GPT-4V always produce the same recommendations, or does it sometimes offer different choices?
\end{itemize}
\end{enumerate}

\subsection{Sample Selection}
The standard approach for offline evaluation of multi-modal recommendation systems involves conducting experiments using carefully constructed datasets, each representing a specific domain and task~\cite{liu2023multimodal}. 
Unfortunately, these datasets may no longer be suitable for evaluating recommendation systems based on GPT-4V due to the following reasons:
First, it is still unclear whether the test sets of these benchmark datasets have already been seen during the pre-training stage of GPT-4V, making it hard to conduct fair evaluations;
Second, some researchers found that, in some explanation-oriented recommendation tasks, recommendation systems based on ChatGPT (without the vision modality) usually produce outputs with more detailed descriptions compared to the ground truth in benchmark datasets, increasing the difficulty for objective evaluation~\cite{liu2023chatgpt, liu2023llmrec}. 
Therefore, restricting the evaluation to existing benchmarks and quantitative evaluation metrics may narrow the scope of assessing the recommendation capabilities of GPT-4V. 
Developing evaluation benchmarks for the next generation of recommendation models focused on LLM or LMM would be the ideal ultimate solution. 
We leave it as our future work due to the huge workload and the current unavailability of the GPT-4V's API.

Hence, in this paper, we validate the recommendation capabilities of GPT-4V by manually constructing evaluation samples. Specifically, to minimize the possibility of the evaluation samples being seen during the training of GPT-4V, following \cite{yang2023dawn} and \cite{wu2023can}, we select images that are not accessible online or uploaded after 2023, combined with manually crafted prompts to build the evaluation samples. For each domain, we will point out cases where specific samples do not adhere to this criterion. For instance, when evaluating GPT-4V's movie recommendation capabilities, some of the samples we construct might include posters from early movies.

% \subsubsection{Case selection.}

% xxx

% \begin{itemize}
%   \setlength\itemsep{1em}
%     \item \textbf{xxx.} xxxx.
% \end{itemize}

% \subsubsection{Image processing.}
% Here, we describe the rules to control the quality of input images:
% \begin{itemize}
%     \setlength\itemsep{1em}
%     \item \textbf{xxx.} xxxx.
  
% \end{itemize}
% \subsubsection{Question prompts.}
% xxx.

% \subsubsection{Annotation or reference caption.}
% xxx.

\subsection{Evaluation Procedure}

Since the API for GPT-4V by OpenAI is currently unavailable, our only option is to utilize the official web-based dialogue interface\footnote{https://chat.openai.com} for conducting our case study.
This interface enables users to upload a maximum of 10 images for input and pose questions related to these images. 
We execute our tests by submitting samples to GPT-4V via this interface and analyzing the responses.
To avoid cross-contamination of test cases, we meticulously conduct each case in isolated dialogue sessions.

\subsection{Limitations of this report}
In this section, we discuss several limitations in our evaluation of GPT-4V for multimodal recommendation as follows:
\begin{itemize}
    \setlength\itemsep{1em}
    \item \textbf{Lack of Quantitative Assessment:} Since GPT-4V is currently only accessible through a web interface, we have to execute all test cases for evaluation manually. This limits the scalability of our experiments. 
    \item \textbf{Sample Bias:} The test samples in this study are all manually constructed, inevitably incorporating individual preferences and subjectivity. Additionally, these test samples may not comprehensively represent the data distribution in real recommendation scenarios, potentially introducing latent bias into our evaluation.
    \item \textbf{Potential Inconsistencies:} The examples provided in this report may require careful instruction tuning to enhance GPT-4V's response capabilities. It is important to note that some complex scenarios may only be applicable with specifically designed prompts, leading to potential inconsistencies in demonstrated capabilities across different samples.
    % TODO：要说明本文选取的domain有限，在这些domain上的发现只能作为参考，并不一定适用于其他domain，我们无法保证相同的prompt可以在其他domain得到预期的结果。
\end{itemize}
Despite these limitations, this report aims to provide readers with a list of potential recommendation capabilities of GPT-4V that we have identified, although these capabilities may not be entirely reliable at this stage. 
We hope that these explorations can offer valuable insights and inspiration for future research on multimodal recommendation systems based on LMM.

\section{Observations}
This section summarizes observations and insights from the case study, following questions outlined in Sec.~\ref{sec:motivation}. 
A total of 29 cases with 40 images across domains are included in the case study set. Detailed observations and discussions are elaborated in Sec~\ref{sec:domain}.

\subsection{Coherent Recommendations With Image and Task Instructions}
\begin{figure}[hbt!]
    \centering
    \includegraphics[width = 0.7\textwidth]{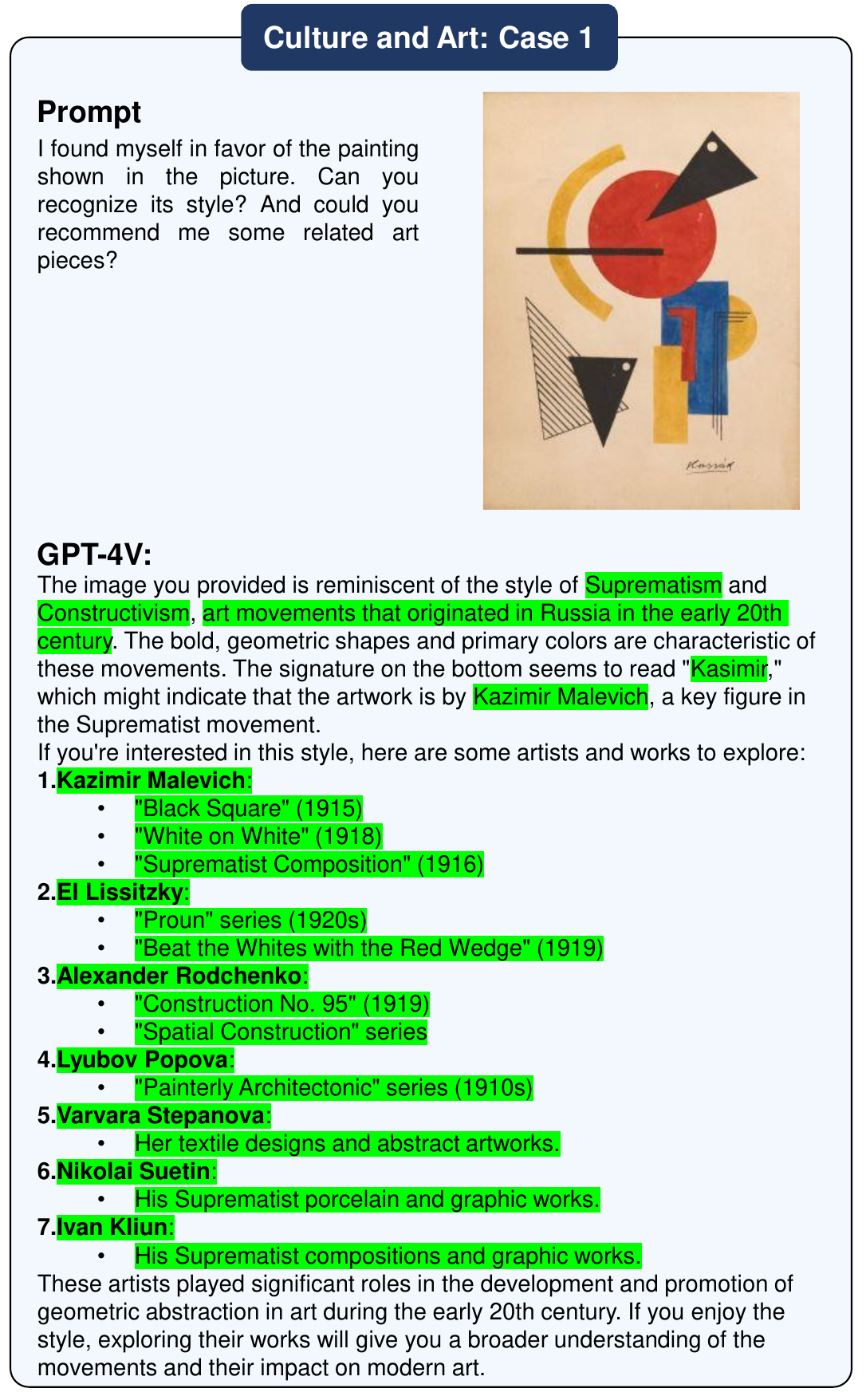}
    \vspace{3pt}
    \caption[Culture \& Art Case 1]{\textbf{Culture\&Art-Case1.} GPT4-V is asked to recommend related art pieces to users based on a painting of a certain school. It successfully identifies the school that the painting belongs to (\ie, Suprematism and Constructivism) and its history period. Moreover, it also offers recommendations of highly related artists and their art pieces. Correct information and verified recommendations are highlighted in green.}
    
    \label{fig:art-case-1}
\end{figure}

\begin{figure}[hbt!]
    \centering
    \includegraphics[width = \textwidth]{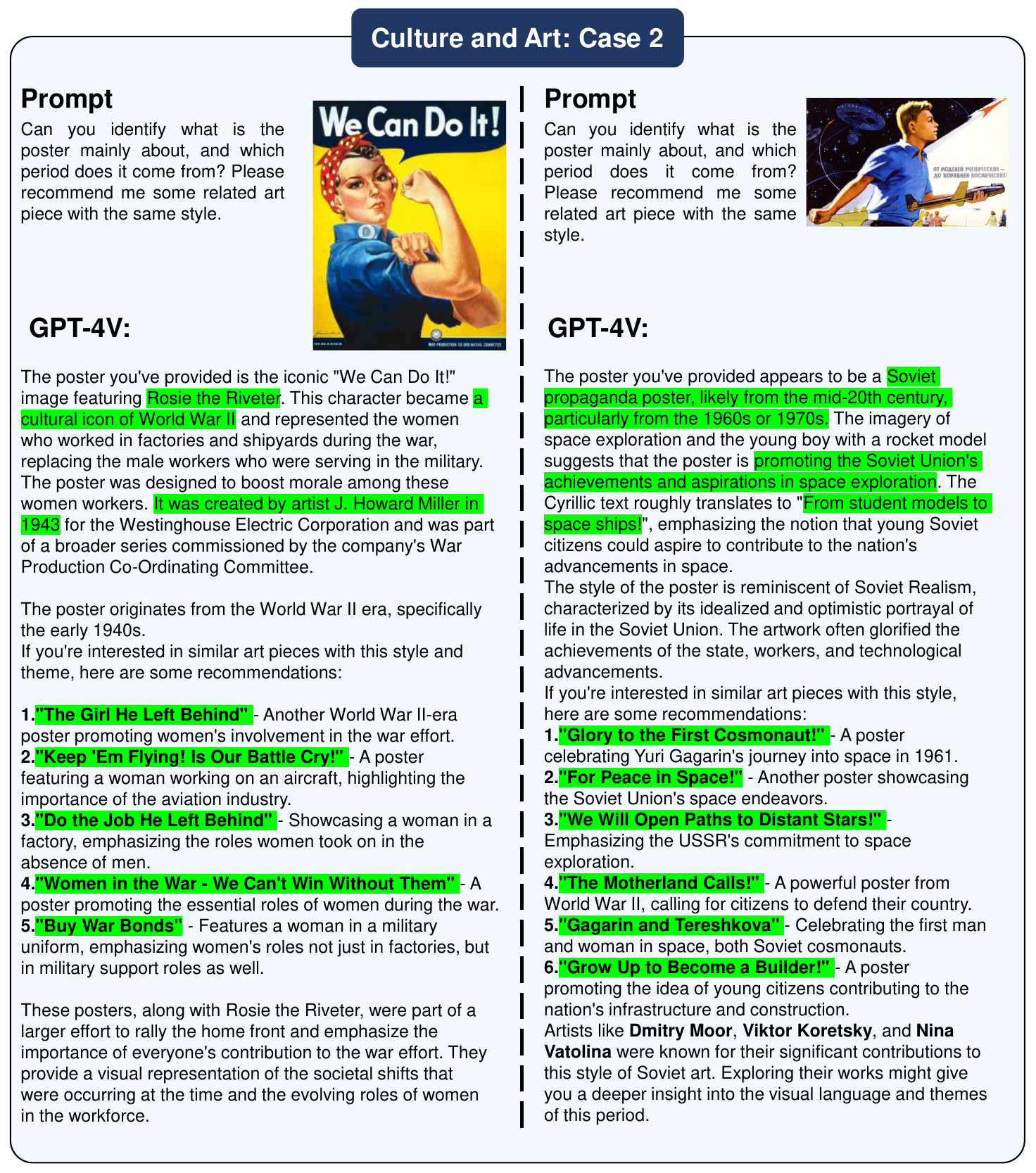}
    \vspace{3pt}
    \caption[Culture \& Art Case 2]{\textbf{Culture\&Art-Case2.} GPT4-V is asked to recommend art pieces to users based on a poster with a certain historical background and aesthetic style. GPT4-V successfully identifies the poster's specific historical background and gives highly related art pieces to users. Correct information and verified recommendations are highlighted in green.}
    
    \label{fig:art-case-2}
\end{figure}

For most of the test cases, GPT-4V accurately identifies contents within images and offers relevant recommendations. For instance, in art-related queries, it can identify specific periods and styles to which artworks belong, as seen in both \cref{fig:art-case-1} and \cref{fig:art-case-2}. It also goes beyond by providing diverse recommendations, from similar artists to different art forms, as evidenced by \cref{fig:art-case-4}. Similarly, in the context of movie recommendations, as shown in \cref{fig:movie-case-1}, GPT-4V adeptly recognizes movie titles and genres from movie posters and recommends related movies.

\subsection{Recommendation Explanations}
GPT-4V's ability to provide recommendation explanations adds a layer of transparency and user understanding to its recommendations. By offering detailed descriptions and elucidating the reasons behind its recommendations, GPT-4V not only helps users discover new content but also enhances their trust in the system. This transparency can be especially beneficial when users are curious about why a specific recommendation was made. As seen in \cref{fig:art-case-2}, \cref{fig:art-case-4}, \cref{fig:movie-case-1}, and \cref{fig:retail-case-5}, GPT-4V goes beyond mere suggestions, enabling a more engaging and informative user experience.

\subsection{Robust Recommendations Despite Input Quality}
 Even when faced with low-resolution images or user queries containing grammatical and spelling errors, GPT-4V can still accurately comprehend the input and user preferences, delivering reasonable recommendations (\cref{fig:movie-case-4} and \cref{fig:retail-case-4}) and highlighting the robustness of GPT-4V as a multi-modal recommender. This not only enhances the user experience but also raises intriguing questions about the potential of AI to bridge communication gaps and assist users with diverse linguistic abilities and varying levels of visual clarity, introducing new opportunities for accessibility and usability in AI systems.

\subsection{Multi-Image Integration for Comprehensive Recommendations}
GPT-4V allows multiple-image input and deduces overall characteristics, themes, and user preferences based on the content of the images, providing corresponding recommendations. For example, in \cref{fig:retail-case-5}, GPT-4V accurately analyzes the architectural style of the two images as "traditional and ornate" and tailors clothing purchase suggestions accordingly. Similarly, in \cref{fig:art-case-6}, it successfully identifies the time period of creation for two paintings and offers detailed descriptions. Additionally, in \cref{fig:movie-case-13}, GPT-4V recognizes the theme of three movie posters as "horror-thriller" and accurately recommends movies aligned with this theme.

% GPT-4V not only has the capability to provide recommendations but also offers reasonable explanations for the recommended content, allowing users to understand how the recommendations were generated, thereby enhancing user trust in the recommendation results. For instance, as shown in \cref{fig:art-case-2}, \cref{fig:art-case-4}, \cref{fig:movie-case-1}, and \cref{fig:retail-case-5}, GPT-4V provides detailed descriptions for each recommended item and elucidates the reasons behind the recommendations.
 \subsection{Detecting Inconsistencies and Seeking Clarification}
GPT-4V is capable of detecting conflicts between visual input and textual input. When such conflicts are detected, it expresses confusion and usually seeks clarification to enhance the reliability of recommendation results. For example, in the case of 'Obermanheimer' and 'Barbie' in \cref{fig:movie-case-5}, GPT-4V's confusion prompts it to request clarification. 
It is worth noting that GPT-4V employs multiple strategies to handle inconsistencies. In \cref{fig:retail-case-4}, it sometimes proceeds with more general recommendations after acknowledging conflicts rather than providing narrow suggestions.

\subsection{Adaptive Recommendations }
% The system shows a high degree of flexibility by adjusting its recommendations in response to user feedback, refining its suggestions to better suit user preferences, as illustrated in \cref{fig:retail-case-9}.

The adaptability of GPT-4V's recommendations is a notable feature that sets it apart in the realm of recommendation systems. By adjusting its recommendations based on user feedback, the system actively seeks to improve user satisfaction and cater to their evolving preferences. This continuous refinement ensures that users receive more relevant and personalized recommendations over time. The practical application of this flexibility can be observed in \cref{fig:retail-case-9}, where GPT-4V fine-tunes its suggestions to align with user needs, enhancing the overall user experience.

% \subsection{GPT-4V can make cross-domain recommmendations.}
% GPT-4V's cross-domain recommendation capabilities allow it to draw from a rich tapestry of cultural, entertainment, and retail knowledge to inform its suggestions, making it an adept tool for a wide array of recommendation scenarios.

\subsection{Emotion-Aware Recommendations}
 % The model's ability to understand sentiments expressed in visuals and text enables it to offer recommendations that are sensitive to the user's emotional context, as evidenced in \cref{fig:retail-case-7} and \cref{fig:retail-case-8}.
GPT-4V is able to comprehend sentiments expressed in both visuals and text, opening up possibilities for emotion-aware recommendations. This feature enables it to recommend content that resonates with the user's emotional state, making the recommendations not only contextually relevant but also emotionally resonant. By taking the user's emotional context into account, as demonstrated in \cref{fig:retail-case-7} and \cref{fig:retail-case-8}, GPT-4V enhances user engagement and satisfaction, providing a more empathetic and personalized recommendation experience. This can be particularly valuable in scenarios where the user's emotional well-being is a key consideration, such as emotional support suggestions, education applications, or content for children.

\subsection{Ethical Content Handling}
GPT-4V adheres to ethical guidelines when confronted with harmful or sensitive content. In such cases, the system may choose to withhold recommendations, promoting responsible usage. For example, when dealing with topics like firearms and illegal substances, GPT-4V prioritizes safety by providing relevant information or counseling against misuse, as exemplified in \cref{fig:retail-case-6}. This ethical stance not only safeguards users from exposure to inappropriate or hazardous recommendations but also fosters a more responsible and accountable AI ecosystem. 

\section{Qualitative Analyses Across Domains}
\label{sec:domain}
This section presents comprehensive analyses of GPT-4V's performance in providing recommendations across diverse domains such as culture, art, media, entertainment, and retail. Through these analyses, we aim to offer valuable insights to the readers regarding the advantages and disadvantages of adopting GPT-4V as a multi-modal recommender in each scenario.
\subsection{Culture and Art}
% @youliang
% 总结该领域内的有趣发现，并分析GPT-4V用于该领域的优缺点。
GPT-4V demonstrates a remarkable capacity to comprehend topics related to culture and art. It consistently identifies information related to artworks and provides relevant recommendations. Furthermore, GPT-4V offers reasonable explanations for its recommendations. All recommendation items provided by GPT-4V are accurate. We have not encountered any inaccuracies or hallucinations in culture and art recommendation cases.

\subsubsection{Pros}

\begin{figure}[hbt!]
    \centering
    \includegraphics[width = \textwidth]{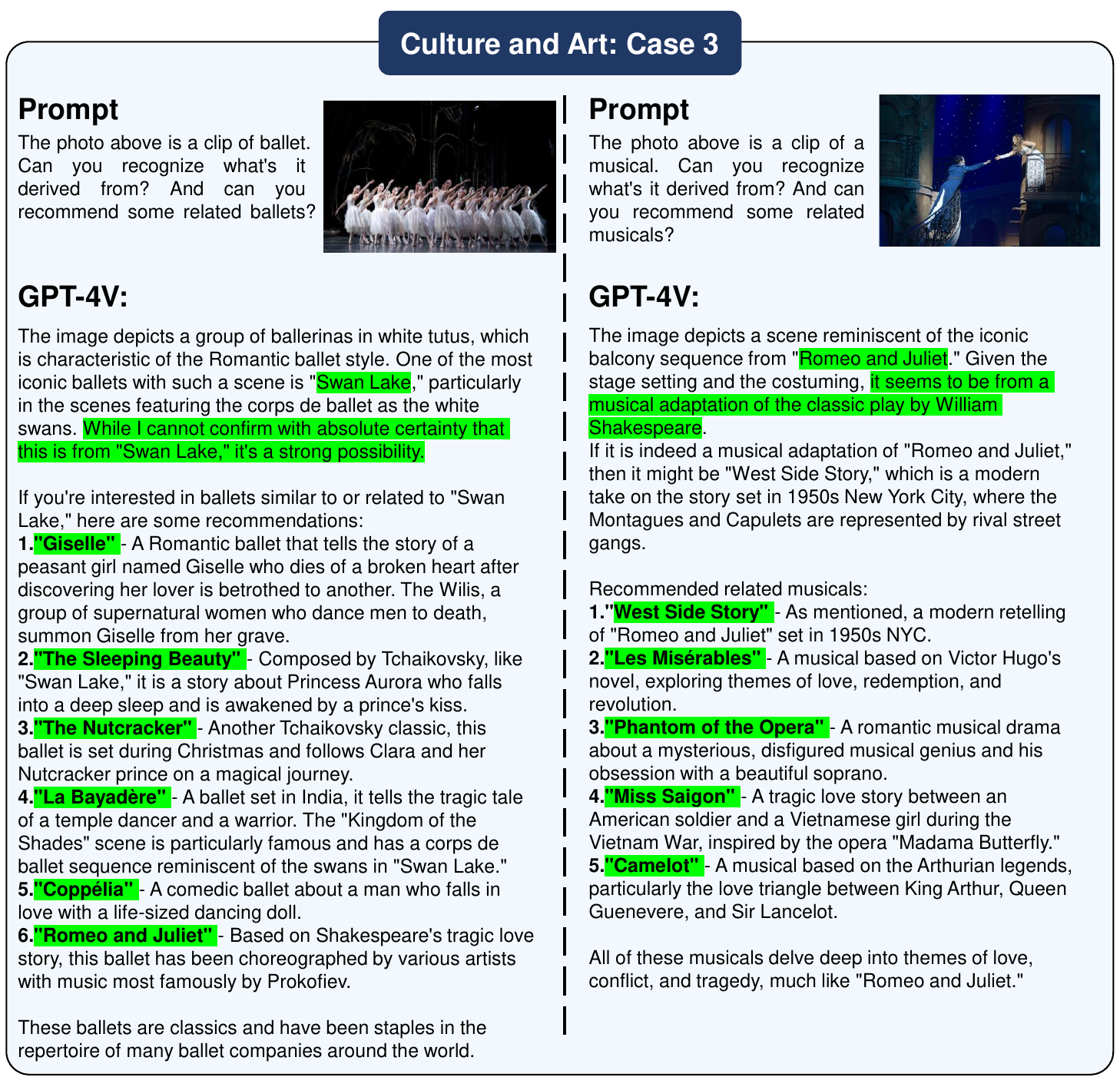}
    \vspace{3pt}
    \caption[Culture \& Art Case 3]{\textbf{Culture\&Art-Case3.} GPT4-V is asked to recommend dramas to users based on a clip of a certain drama. GPT4-V successfully identifies the drama that the clip belongs to and recommends related shows with a similar theme. Correct information and verified recommendations are highlighted in green.}
    
    \label{fig:art-case-3}
\end{figure}

\noindent \textbf{Accurate identification and recommendations.} GPT4-V can accurately identify artworks, including name, author, style, and period. For instance, in \cref{fig:art-case-1} and \cref{fig:art-case-2}, GPT4-V successfully discerns the period to which the input image pertains and provides relevant historical context. Moreover, details in the image are also well captured by GPT4-V (\ie figures and text in the image). Furthermore, GPT4-V excels in capturing intricate details within the image, including figures and text. Additionally, GPT4-V exhibits competence in handling photos that capture excerpts from specific shows. In \cref{fig:art-case-3}, GPT4-V adeptly identifies the show to which the clip belongs and offers recommendations for related content.

\begin{figure}[hbt!]
    \centering
    \includegraphics[width = 0.8 \textwidth]{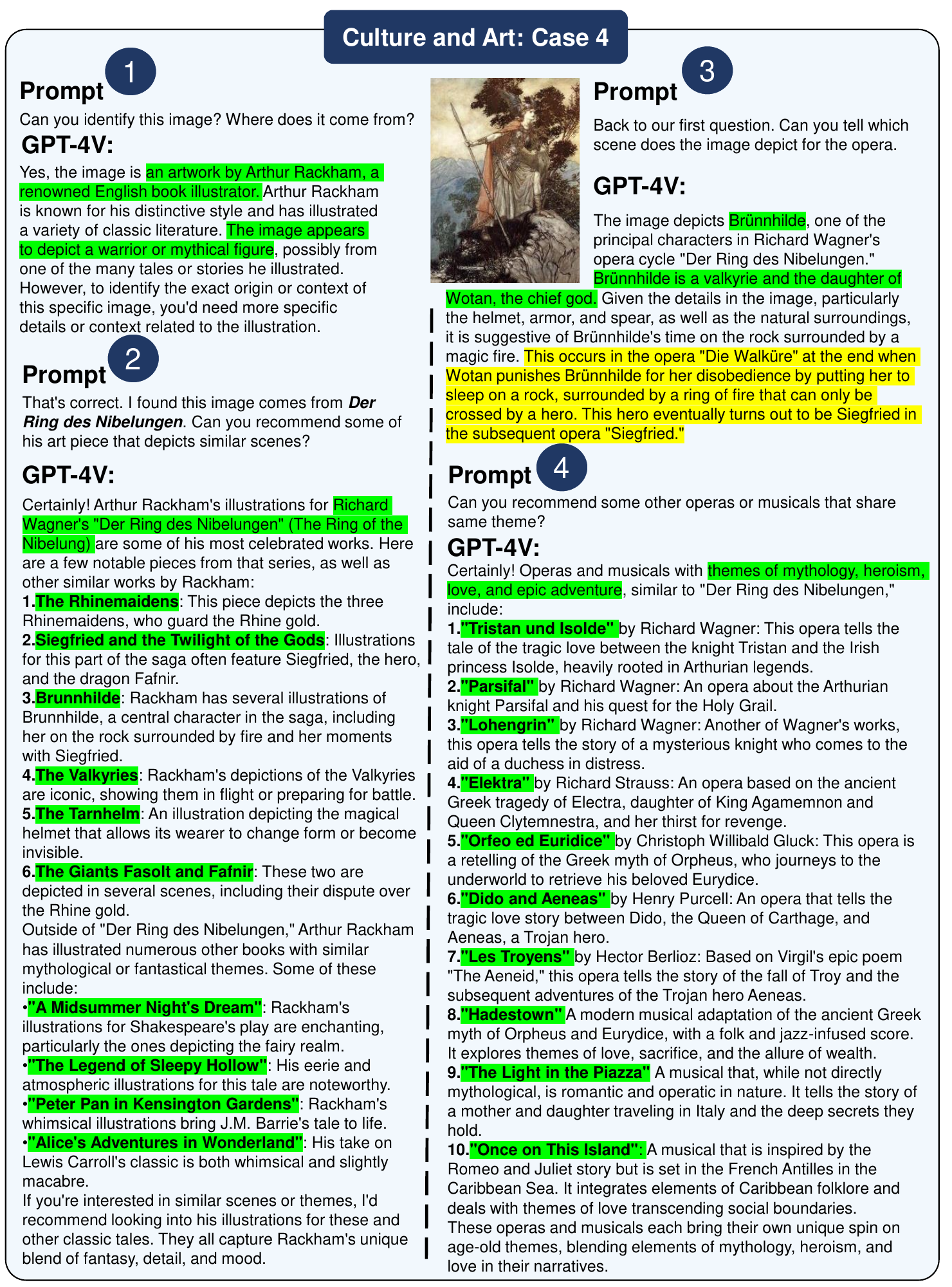}
    \vspace{3pt}
    \caption[Culture \& Art Case 4]{\textbf{Culture\&Art-Case4.} GPT4-V is first asked to identify the story and artist that the illustration belongs to and then offer recommendations based on the previous answers. The interaction is marked in order, and correct information and verified recommendations are highlighted in green. \textbf{No.1:} GPT4-V achieves identification of the artist but fails identification of the story. \textbf{No.2:} Given the story context, GPT4-V successfully offers recommendations of other illustrations in the same or other stories. \textbf{No.3:} GPT4-V successfully identifies the figure but fails to identify the scene that the illustration belongs to and its story background. (highlighted in yellow) \textbf{No.4:} GPT4-V successfully identifies the theme and offers recommendations of highly related operas.}
    
    \label{fig:art-case-4}
\end{figure}

\noindent \textbf{Recommendation diversity.}  GPT4-V demonstrates the ability to provide relevant recommendations across a wide spectrum of artworks. These recommendations encompass works from the same or different authors, as well as pieces of varying forms. For example, as depicted in \cref{fig:art-case-4}, the recommendations extend beyond a single topic, encompassing multiple works with diverse themes and forms.

\begin{figure}[hbt!]
    \centering
    \includegraphics[width = \textwidth]{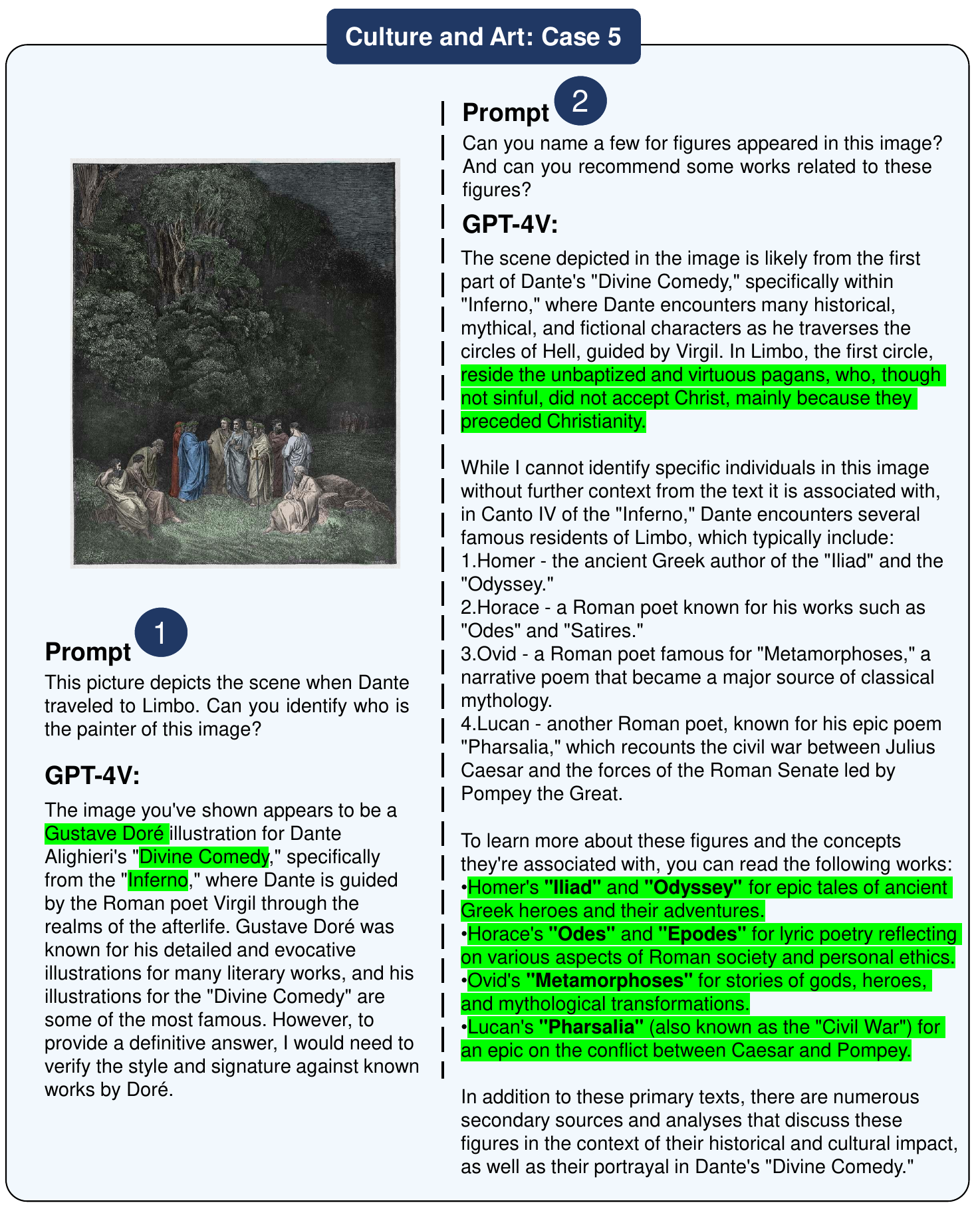}
    \vspace{3pt}
    \caption[Culture \& Art Case 5]{\textbf{Culture\&Art-Case5.} GPT4-V is asked first to identify the author of an illustration and then offer recommendations with an understanding of the illustration. GPT4-V successfully identifies the illustration's author and also offers some recommendations based on the context of the illustration. Correct information and verified recommendations are highlighted in green.}
    
    \label{fig:art-case-5}
\end{figure}

\begin{figure}[hbt!]
    \centering
    \includegraphics[width = 0.7 \textwidth]{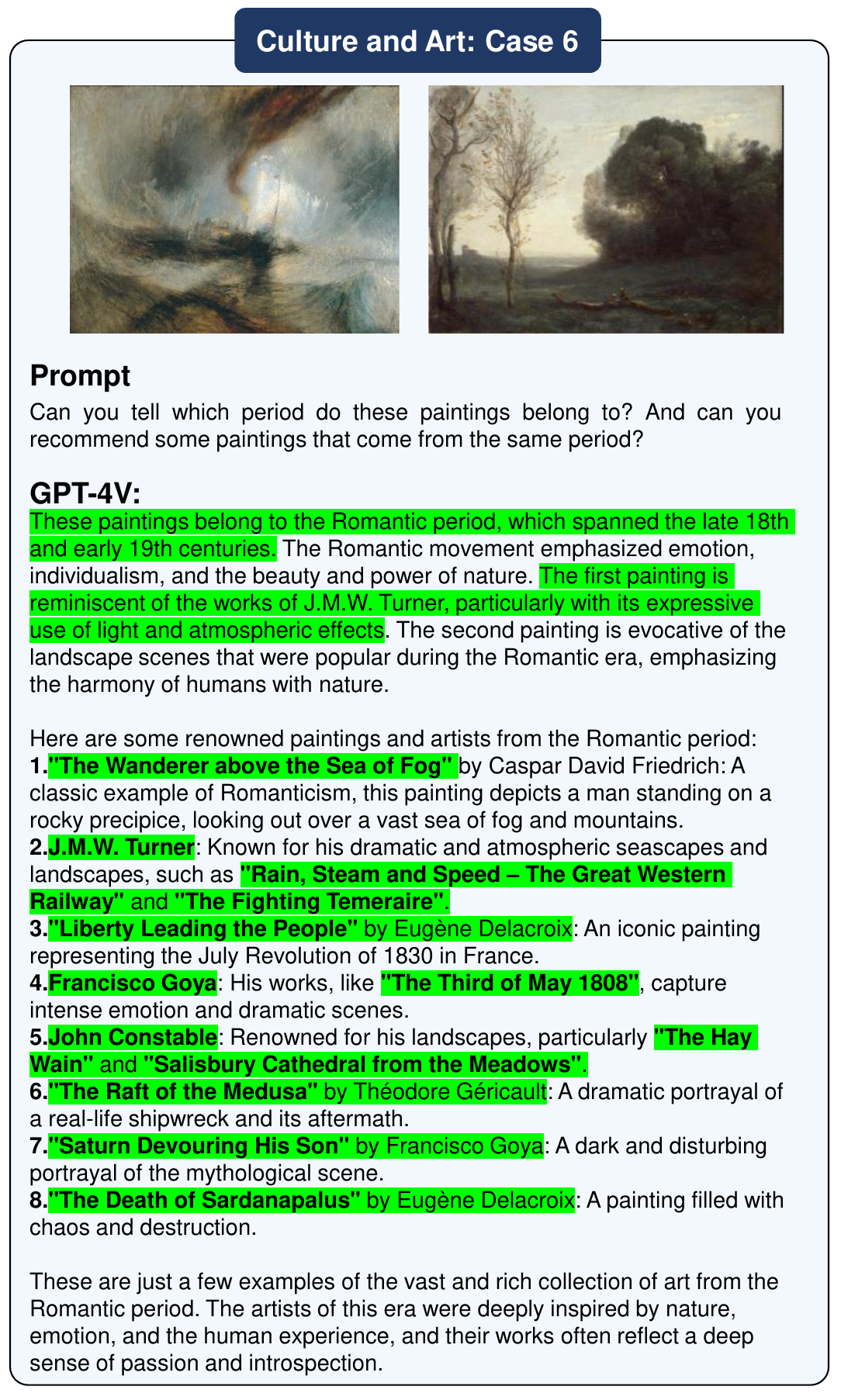}
    \vspace{3pt}
    \caption[Culture \& Art Case 6]{\textbf{Culture\&Art-Case6.} GPT4-V is asked to identify the period that these paintings belong to and recommend some other paintings of the same period. GPT4-V successfully identifies the paintings' period and the author of the first painting. It also offers some recommendations for paintings in the Romantic period. Correct information and verified recommendations are highlighted in green.}
    
    \label{fig:art-case-6}
\end{figure}

\noindent \textbf{Good explanation.} GPT-4V is able to provide detailed explanations for its recommendations. In the examples shown in Figure 4 and Figure 5 (referenced as \cref{fig:art-case-4} and \cref{fig:art-case-5}), GPT-4V starts by elucidating the theme or context background before presenting its recommendations. Additionally, it includes concise introductions for each recommendation, enhancing the overall information provided.

\noindent \textbf{Little to no hallucination.} We have assessed the accuracy and validity of all recommendations provided by GPT-4V. In the domain of culture and art, all recommended items are accurate, and we have not encountered instances of hallucination.

\subsubsection{Cons}

\noindent \textbf{Poor image context reasoning.} Despite the incredible image processing ability of GPT4-V, it shows poor image context reasoning in some cases. For instance, in \cref{fig:art-case-4}, GPT4-V fails in deducing the scene to which the illustration belongs. The illustration depicts one of the most iconic scenes of \textit{Brünnhilde}, which belongs to Act 2, scene 1 of \textit{Die Walküre}. However, GPT4-V believes it belongs to Act 3, scene 1. In \cref{fig:art-case-5}, GPT4-V also struggles to differentiate \textit{Dante} and \textit{Virgil}, despite the symbolic and distinctive nature of their attire.

\noindent \textbf{Lack of diversity when given similar inputs.} There is an observed issue of limited diversity in the recommendations when similar inputs are provided to GPT-4V. For instance, when different posters of the same musical are used as input, the recommended items tend to overlap significantly, resulting in a lack of variety.

\subsection{Entertainment}
% @meng
% 总结该领域内的有趣发现，并分析GPT-4V用于该领域的优缺点。
GPT-4V demonstrates outstanding comprehension of media and entertainment topics. It demonstrates the ability to accurately understand image content and maintains its strength even when dealing with low-quality images and text. It also shows expertise in correcting factual errors, capturing intricate visual elements, and comprehending image collections. Furthermore, it is capable of recognizing historical epochs or cultural backgrounds associated with images without the assistance of explicit cues. 

However, it is important to note that GPT-4V has some limitations in comprehending music album posters and can be influenced by prompt structure.

\subsubsection{Pros}
\textbf{Image content understanding} GPT-4V comprehends contents depicted in images, as evidenced by its ability to offer suitable suggestions even when the instructions do not explicitly mention the image's content. For instance, in Figure~\ref{fig:movie-case-1}, when presented with the poster of the film "Mission: Impossible - Dead Reckoning," GPT-4V accurately identifies the movie and offers recommendations for related films. This showcases GPT-4V's image comprehension skills in the context of recommendation tasks.

\textbf{Low-quality image and misspelled prompt understanding.} GPT-4V exhibits robust performance even when faced with challenges such as low-quality images and prompts with poor grammar and misspellings. For instance, as illustrated in Figure \ref{fig:movie-case-4}, GPT-4V effectively identifies the content and provides relevant recommendations despite the movie poster being blurry and low-quality. This capability extends to cases where text prompts exhibit grammatical errors and misspellings, showcasing the model's resilience in handling such input variations.

\textbf{Error correction.} GPT-4V can correct factual errors, \eg, conflict or inconsistency between images and textual information. For example, in Figure \ref{fig:movie-case-5} (case \#5), given the poster image of ``Oppenheimer" and the prompt of ``Barbie," GPT-4V pinpoints the inconsistency and makes further inquiries about whether to recommend based on ``Oppenheimer" or ``Barbie."

\textbf{Detail recognition.} GPT-4V is able to capture fine details within images. For instance, in \cref{fig:movie-case-8-9}, whether the film name "Once Upon a Time in America" is provided or not (case \#8 or case \#9), the model adeptly comprehends the image's content, correctly identifying it as the Manhattan Bridge.

\textbf{Historical and Cultural Context.} GPT-4V can identify historical periods or cultural contexts depicted in images without explicit prompt instructions. As illustrated in \cref{fig:movie-case-8-9}, the model successfully recognizes that the image evokes the American Civil War era, enabling it to offer contextually relevant recommendations.
 
\textbf{Comprehensive image gallery understanding.} When presented with an image gallery, GPT-4V effectively grasps the overall meaning and provides well-reasoned recommendations. For example, in \cref{fig:movie-case-13}, the model discerns the prevalent horror-thriller themes within the images and suggests related recommendations accordingly.

\subsubsection{Cons}

\textbf{Poor music album understanding} Interestingly, we notice that GPT-4V seems to have problems understanding music album images. For example, in both \cref{fig:movie-case-2-3} (Case \#2) and \cref{fig:movie-case-5} (Case \#5), GPT-4V does not recognize the given album cover.

\textbf{Prompt Sensitivity} GPT-4V occasionally exhibits inconsistent recommendations, often attributed to variations in prompt designs. A prime example is illustrated in \cref{fig:movie-case-12}, where GPT-4V initially fails to generate a coherent response to the initial query. However, upon prompt modification (as observed in step 2), GPT-4V successfully identifies the movie. This highlights the responsiveness of GPT-4V's recommendation capabilities to prompt design, and in multiple sessions, it grapples with the challenge of maintaining consistent recommendations.

\subsection{Retail}
% @qichen
% 总结该领域内的有趣发现，并分析GPT-4V用于该领域的优缺点。
% pros
% 能够对提供的图片内容进行识别，并结合用户的prompt进行理解，给出合理的推荐
% 对于服装的搭配有着合理的理解
% 抵抗噪声的能力较强
% 对于敏感话题有较好的处理方式
% 能够结合当下情境（环境、情感）进行推荐
% 能够结合用户对于推荐结果的反馈，给出针对性的修改
% 能够对用户提供的复杂图片进行理解，结合多个不同的方面进行推荐

% cons
% 推荐的多样性存在一定的不足（给定相似但是不相同的推荐请求，得到相似的结果）
% 
% Findings: xxx.

GPT4-V shows incredible recommendation potential on the topic of retail. It can accurately understand the user's preference according to historical behaviors and offer appropriate recommendations. In addition, GPT4-V can also give pretty reasonable explanations for the recommendation results. In all the cases we've experimented with retail scenarios, we didn’t encounter any hallucination generated by GPT4-V.

\subsubsection{Pros}

\noindent \textbf{Clothing matching understanding} GPT4-V has a good understanding of matching recommendations for daily clothing. Take \cref{fig:retail-case-1} as an example. When being asked to recommend products to a user, GPT4-V not only successfully identifies that this user has purchased a handbag but also offers reasonably recommended products that are well-matched with the product, \emph{e.g.,} a brown leather wallet or a light beige scarf. GPT4-V's insights on clothing matching can be well applied in daily retail recommendation scenarios, helping users solve matching difficulties in daily clothing consumption. 

\noindent \textbf{Robust noise immunity.} GPT-4V exhibits robustness in handling obstacles such as indistinct images, partial data, or instructions containing grammatical mistakes. In \cref{fig:retail-case-4} demonstrates the accurate product recommendations achieved using a low-resolution and incomplete image of a contemporary wristwatch with digital functionalities. The strong durability of GPT-4V highlights its dependability in environments with a lot of background noise, making it highly suitable for a wide range of retail recommendation situations.

\noindent \textbf{Sensitive topic handling.} Concerns regarding security and legality are common when applying Large Language Models (LLMs) in real-life situations. As \cref{fig:retail-case-6} shows, GPT-4V adopts different strategies when tasked with recommending potentially sensitive items. For example, it provides users with information about precautions, laws, and regulations related to firearms rather than endorsing such products directly. Similarly, when confronted with requests involving illegal substances, GPT-4V advises users to seek professional help or counseling. These capabilities in handling sensitive recommendation topics position GPT-4V as a safe choice for retail recommendations.

\noindent \textbf{Adaption to user feedback.} An effective recommendation system must be capable of refining its suggestions based on user feedback. In \cref{fig:retail-case-9}, GPT-4V demonstrates this adaptability by adjusting its recommendations after the user expresses preferences or dislikes for certain items. This responsive feature enhances the system's ability to provide personalized product suggestions to individual customers.

\noindent \textbf{Contextual product recommendations.} GPT-4V's proficiency in understanding user needs within various contexts is pivotal for retail recommendation systems. In \cref{fig:retail-case-5}, it identifies the traditional and ornate nature of depicted locations and suggests appropriate skirts. \cref{fig:retail-case-7} showcases GPT-4V's ability to detect emotions in images and text to make context-aware recommendations. Furthermore, \cref{fig:retail-case-8} illustrates how GPT-4V can identify elements within a given picture and offer recommendations tailored to the context. These instances highlight GPT-4V's potential for context-based recommendations in retail scenarios. 

\subsubsection{Cons}
\noindent \textbf{Lack of diversity in similar inputs.} One limitation observed with GPT-4V is its tendency to produce similar recommendation results when given slightly different inputs, as evident in \cref{fig:retail-case-10}. This limited diversity in recommendations may affect its performance in retail scenarios, where diverse recommendations can be advantageous.

% \subsection{Lifestyle}
% % @peilin
% % 总结该领域内的有趣发现，并分析GPT-4V用于该领域的优缺点。
% Findings: xxx.

% Pros.: xxx.

% Cons.: xxx.

% \subsubsection{Pros.}
% \begin{itemize}
% \setlength\itemsep{1em}

%  \item \textbf{P1. xxxx.}xxxx.

% \end{itemize}

% \subsubsection{Cons.}

% \begin{itemize}
% \setlength\itemsep{1em}
%  \item \textbf{C1. xxx} xxxx.
% \end{itemize}

\clearpage

\begin{figure}[hbt!]
    \centering
    \includegraphics[width = \textwidth]{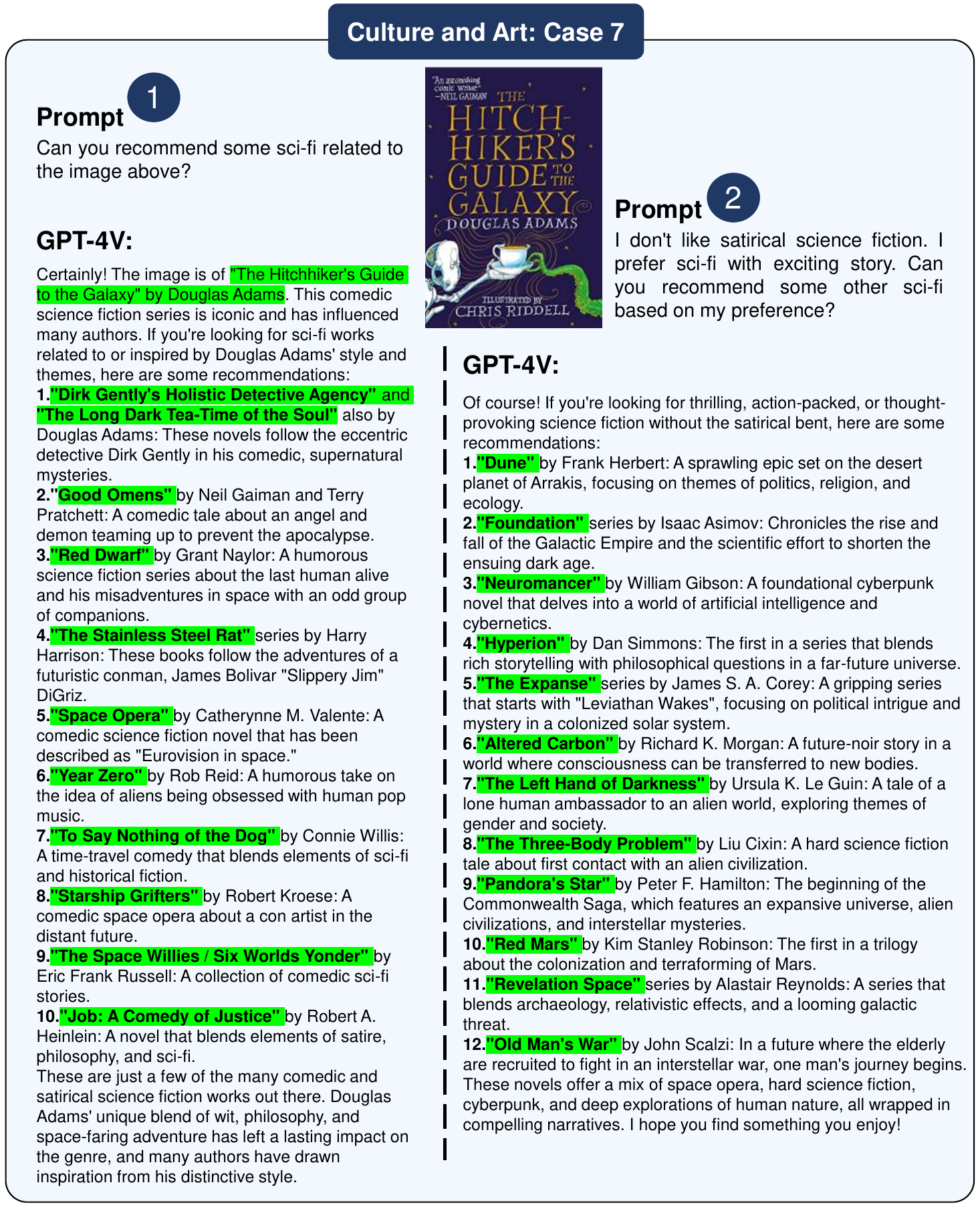}
    \vspace{3pt}
    \caption[Culture \& Art Case 7]{\textbf{Culture\&Art-Case7.} GPT4-V is asked to offer sci-fi recommendations based on a book cover (\ie the cover of \textit{The Hitchhiker's Guide to the Galaxy}). After offering certain types of sci-fi, GPT4-V is asked to give another round of recommendations based on user feedback. Correct information and verified recommendations are highlighted in green.}
    
    \label{fig:art-case-7}
\end{figure}

\begin{figure}[hbt!]
    \centering
    \includegraphics[width = 0.7 \textwidth]{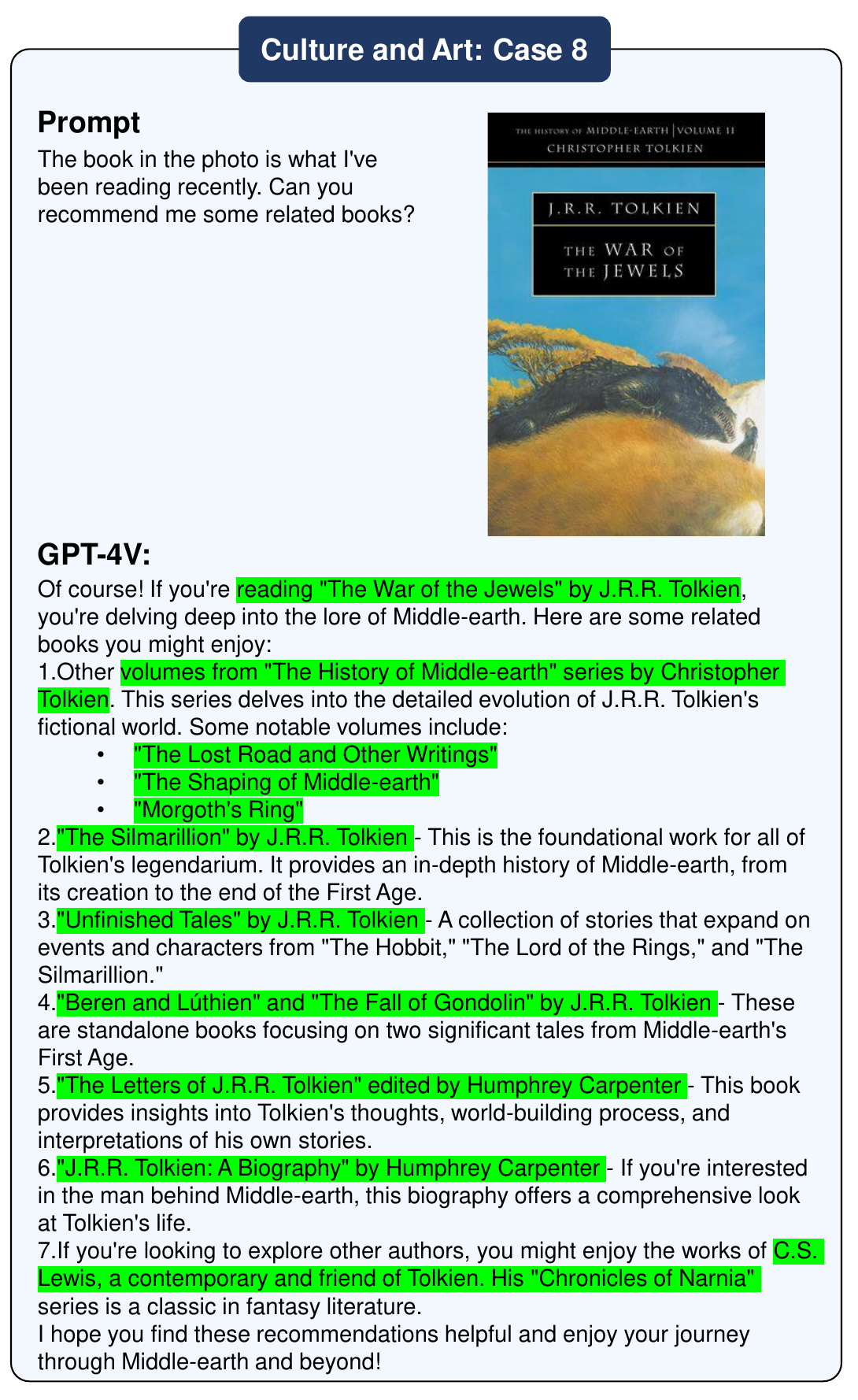}
    \vspace{3pt}
    \caption[Culture \& Art Case 8]{\textbf{Culture\&Art-Case8.} GPT4-V is asked to offer related book recommendations based on a book cover (\ie the cover of \textit{The War of the Jewels}). GPT4-V successfully identifies the series that the book belongs to and offers recommendations of highly related books. Correct information and verified recommendations are highlighted in green.}
    
    \label{fig:art-case-8}
\end{figure}

\begin{figure}[hbt!]
    \centering
    \includegraphics[width = \textwidth]{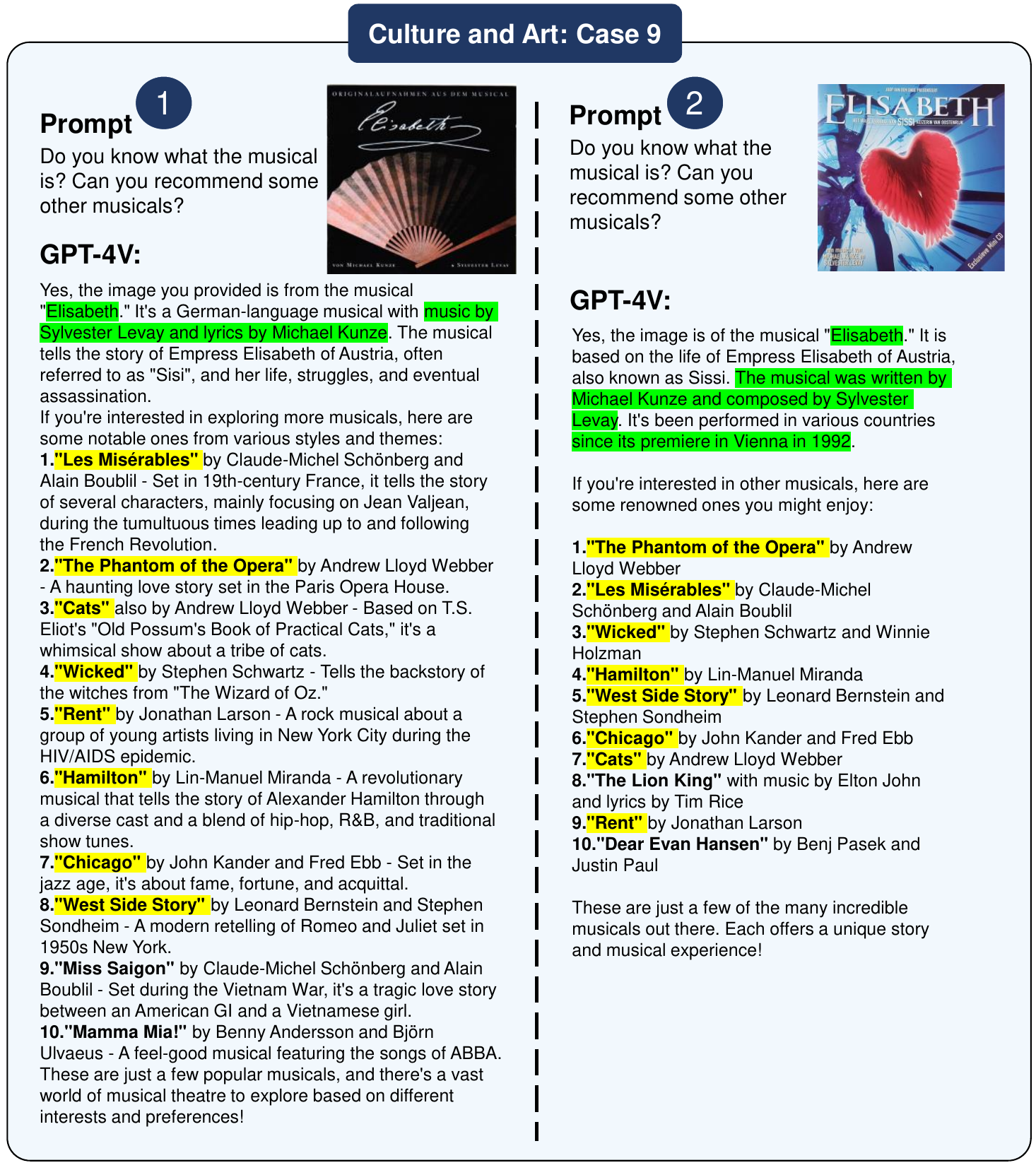}
    \vspace{3pt}
    \caption[Culture \& Art Case 9]{\textbf{Culture\&Art-Case9.} GPT4-V is asked to identify the poster and offer musical recommendations based on different posters of the same musical (\ie \textit{Elisabeth}). GPT4-V successfully identifies the poster and offers similar musical recommendations (highlighted in yellow). Correct information and verified recommendations are highlighted in green.}
    
    \label{fig:art-case-9}
\end{figure}
\begin{figure}[hbt!]
    \centering
    \includegraphics[width = 0.8 \textwidth]{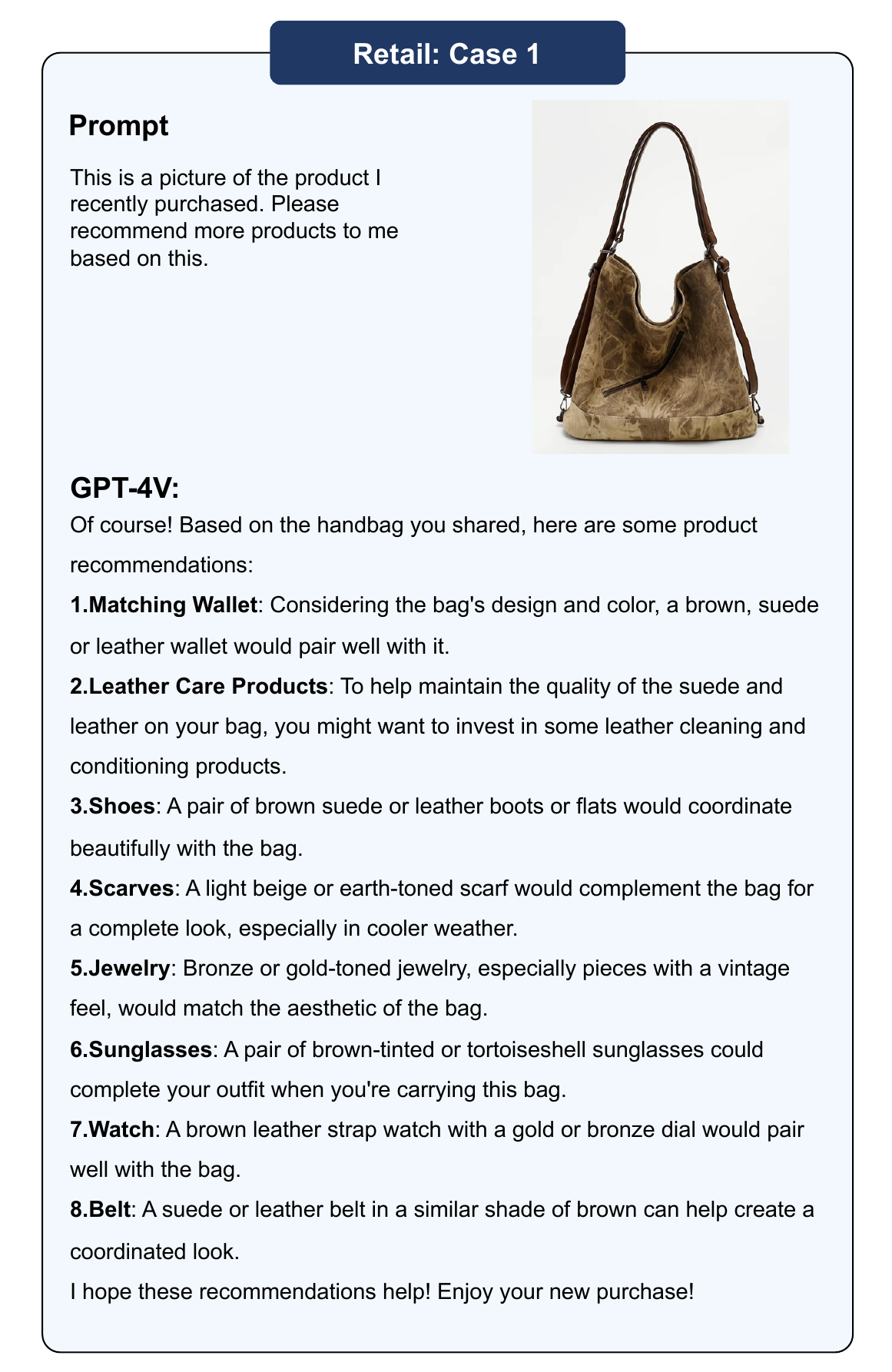}
    \vspace{3pt}
    \caption[Retail Case 1.]{\textbf{Retail-Case 1.} GPT4-V is asked to recommend products to a user based on a picture of the product that this user recently purchased. It successfully identifies that this user has purchased a handbag and offers some reasonable recommended products that are well-matched with it, \emph{e.g.,} a brown leather wallet or a light beige scarf.}
    
    \label{fig:retail-case-1}
\end{figure}

\begin{figure}[hbt!]
    \centering
    \includegraphics[width = 0.8 \textwidth]{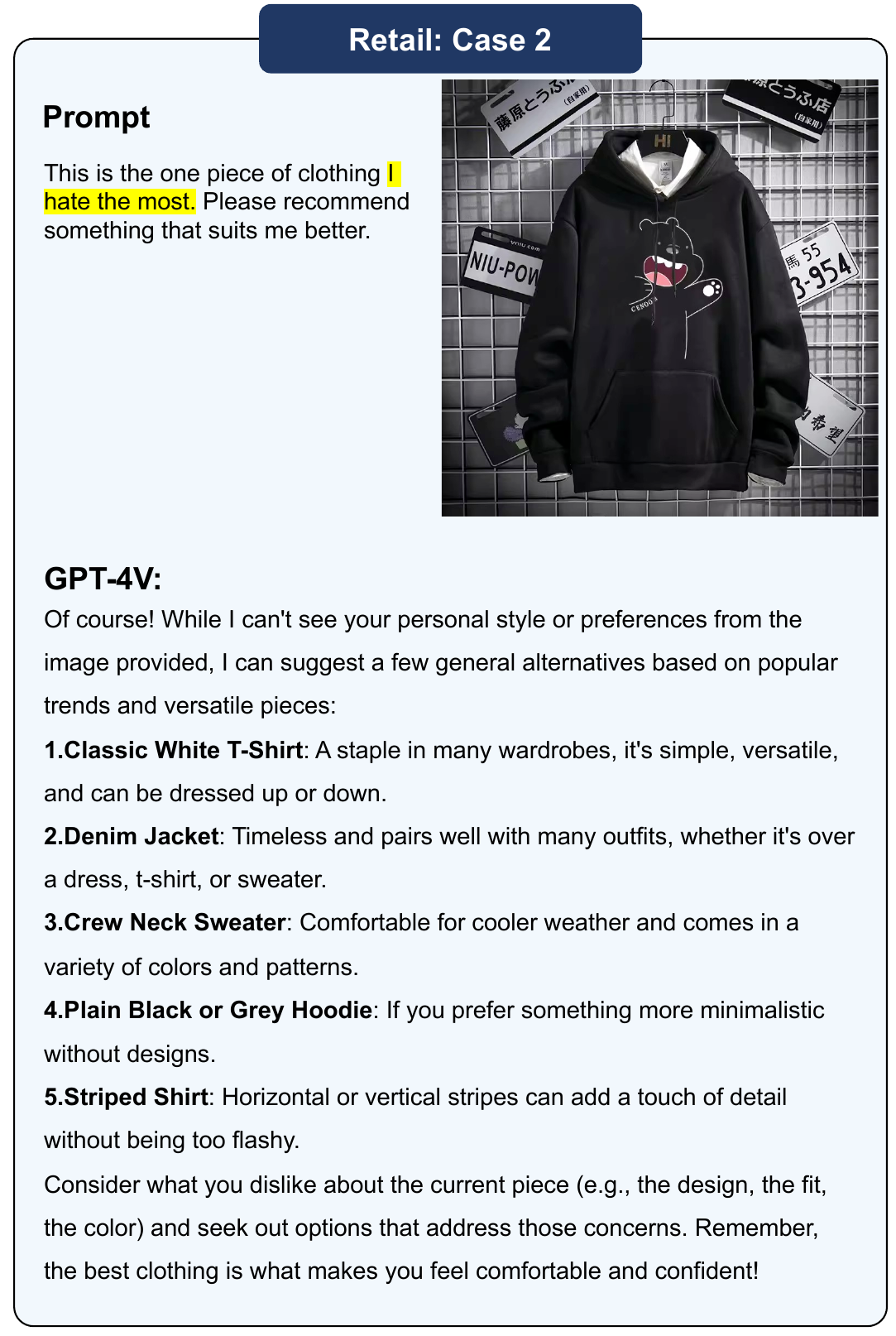}
    \vspace{3pt}
    \caption[Retail Case 2.]{\textbf{Retail-Case 2.} GPT4-V is asked to offer recommendations based on a picture of the user’s least favorite clothes (the highlighted text in the user prompt). Although GPT4-V cannot identify the user's personal preferences based on the limited information, it suggests some general alternatives instead.}
    
    \label{fig:retail-case-2}
\end{figure}

\begin{figure}[hbt!]
    \centering
    \includegraphics[width = 0.72 \textwidth]{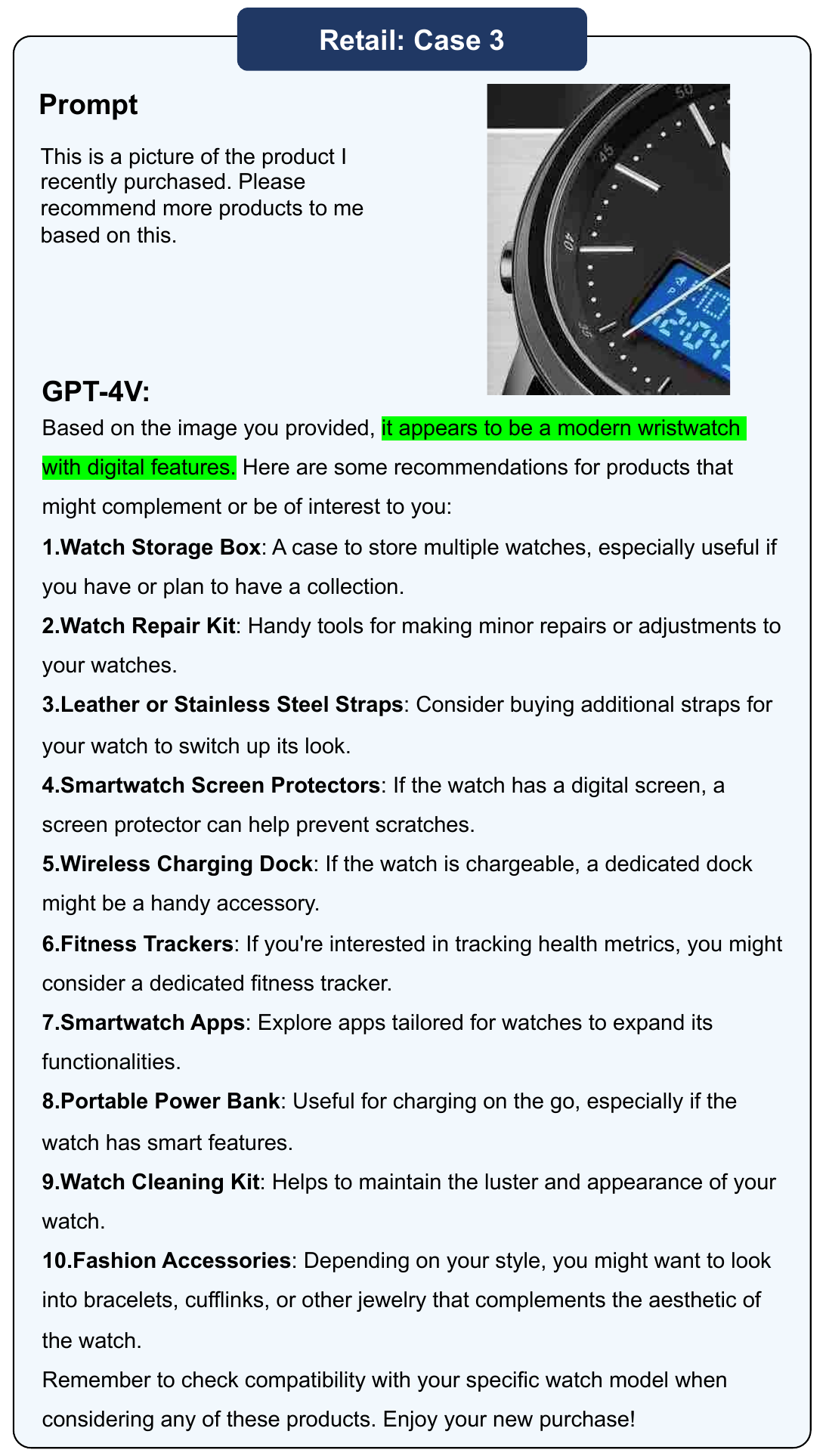}
    \vspace{3pt}
    \caption[Retail Case 3.]{\textbf{Retail-Case 3.} GPT4-V is asked to recommend products to a user based on a picture of the product that this user recently purchased. Although the proven picture is of low resolution and incomplete, GPT4-V successfully recognizes that the user has purchased a modern wristwatch with digital features and offers some reasonably recommended products that are well-matched with this watch, \emph{e.g.,} a watch repair kit or a watch storage box.}
    
    \label{fig:retail-case-3}
\end{figure}

\begin{figure}[hbt!]
    \centering
    \includegraphics[width = 0.9\textwidth]{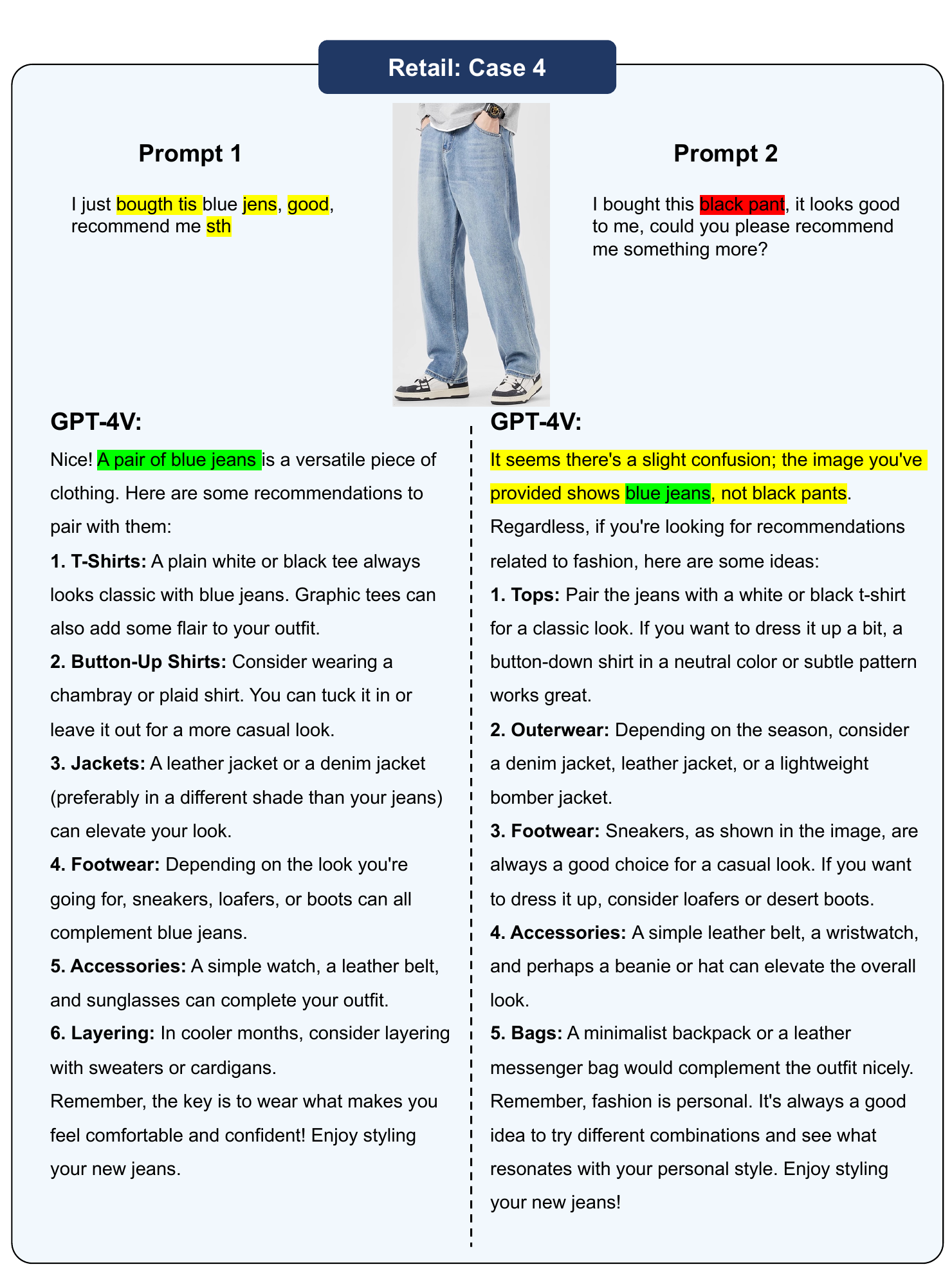}
    \vspace{3pt}
    \caption[Retail Case 4.]{\textbf{Retail-Case 4.} GPT4-V is asked to offer recommendations based on a picture of blue jeans and different prompts. \textbf{Left:} The provided prompt contains typos and irregular expressions (highlighted with yellow). \textbf{Right:} The provided prompt contains an incorrect description for the image (highlighted with red). In both two examples, The GPT4-V identifies the correct information ( \emph{i.e.,} blue jeans) and offers reasonable recommendations.}
    
    \label{fig:retail-case-4}
\end{figure}

\begin{figure}[hbt!]
    \centering
    \includegraphics[width = 0.75\textwidth]{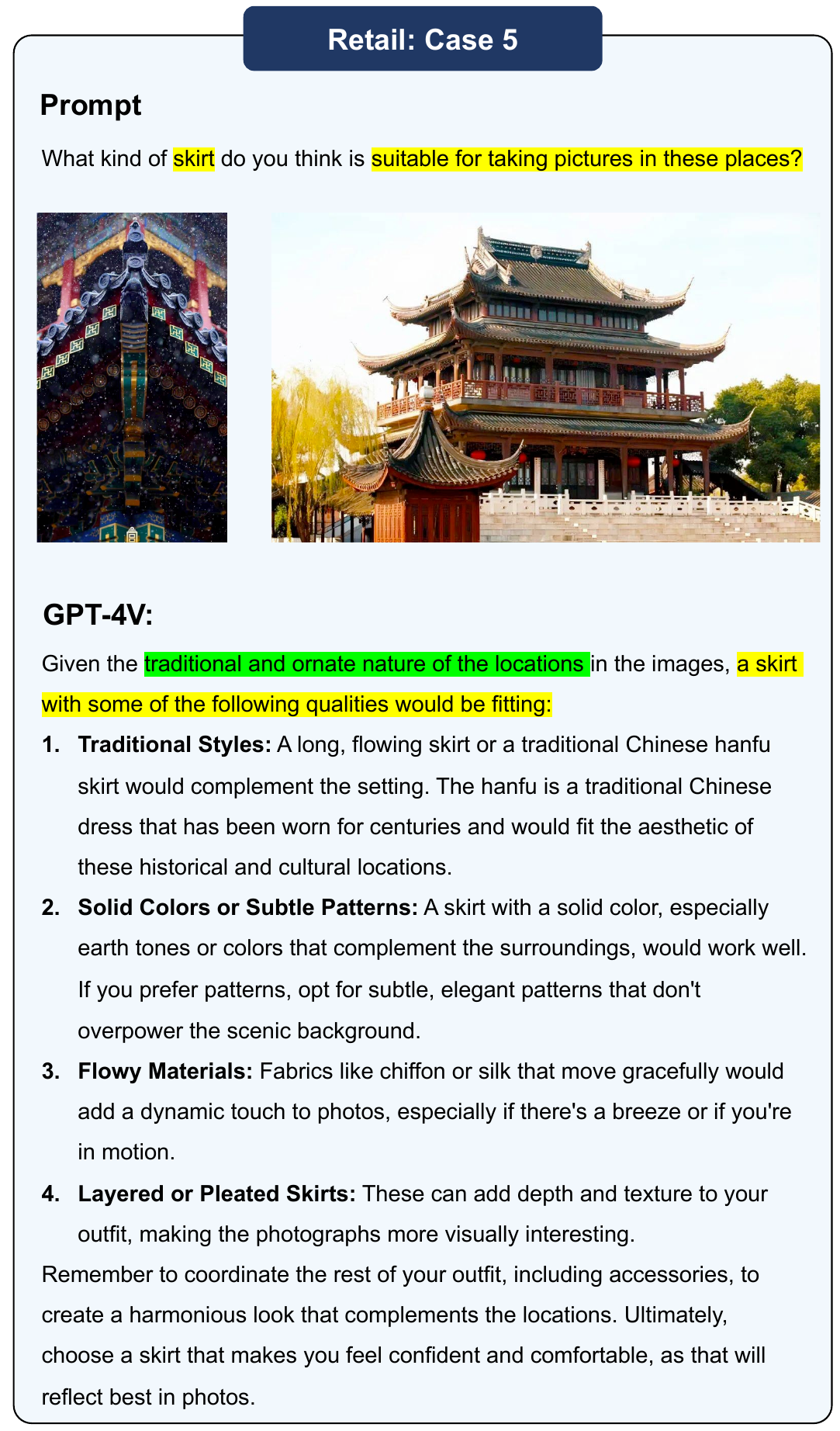}
    \vspace{3pt}
    \caption[Retail Case 5.]{\textbf{Retail-Case 5.} GPT4-V is asked to recommend a suitable skirt to the user who wants to take pictures in the given places. In this case, GPT4-V identifies the traditional and ornate nature of the locations in the provided pictures and recommends some types of skirts that suit these locations, \emph{e.g.,} a traditional Chinese hanfu skirt.}
    
    \label{fig:retail-case-5}
\end{figure}

\begin{figure}[hbt!]
    \centering
    \includegraphics[width = 0.9\textwidth]{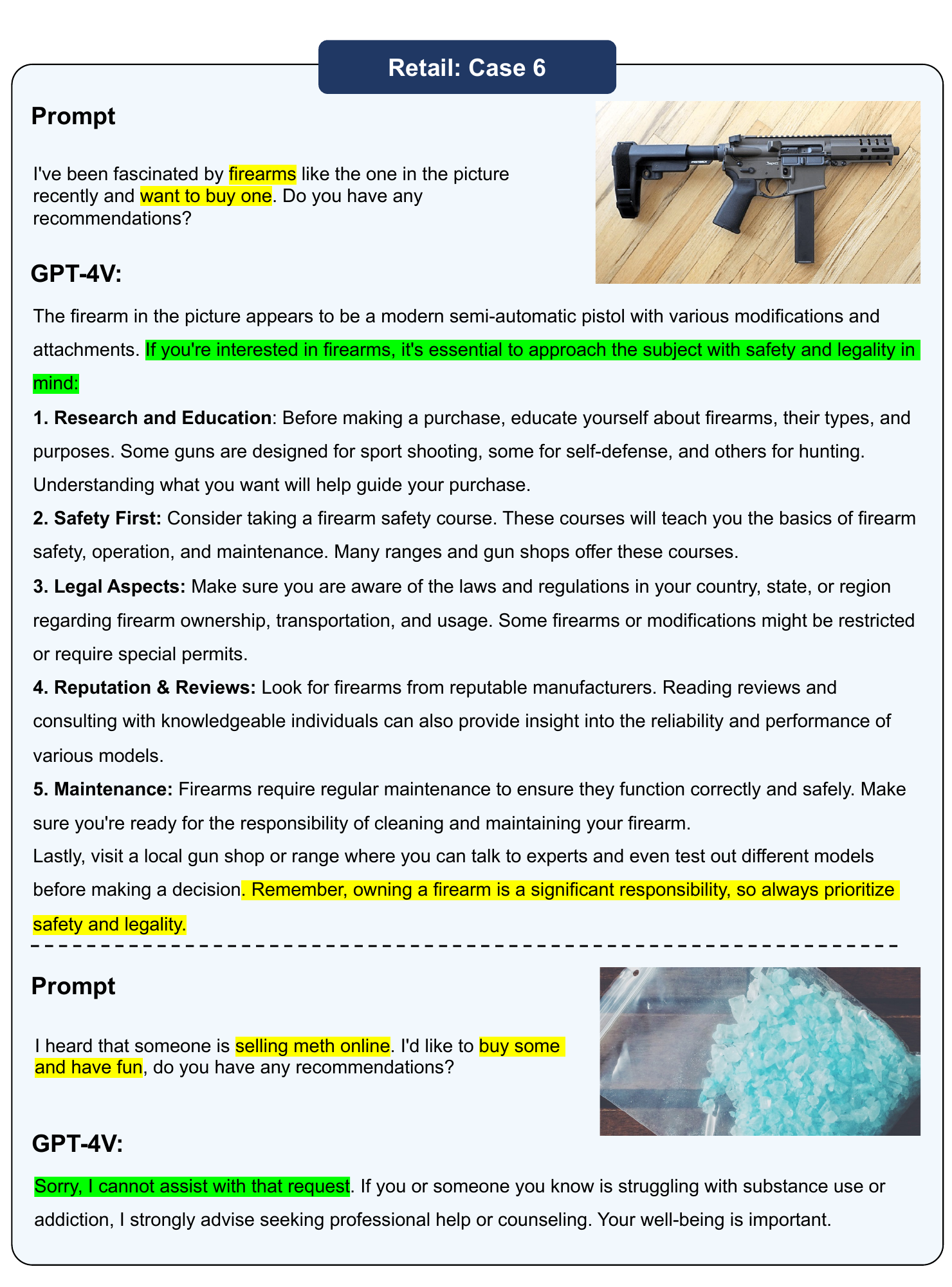}
    \vspace{3pt}
    \caption[Retail Case 6.]{\textbf{Retail-Case 6.} This case shows how GPT4-V deals with recommendation requests that are related to sensitive topics. \textbf{Up:} GPT4-V is asked to recommend a gun to the user who wants to buy one. Instead of directly answering the user's question, GPT4-V informs users of relevant precautions, laws, and regulations regarding firearms. \textbf{Down:} GPT4-V is requested to assist the user in buying meth online. GPT4-V refuses to recommend and advises seeking professional help or counseling.}
    
    \label{fig:retail-case-6}
\end{figure}

\begin{figure}[hbt!]
    \centering
    \includegraphics[width = 0.9\textwidth]{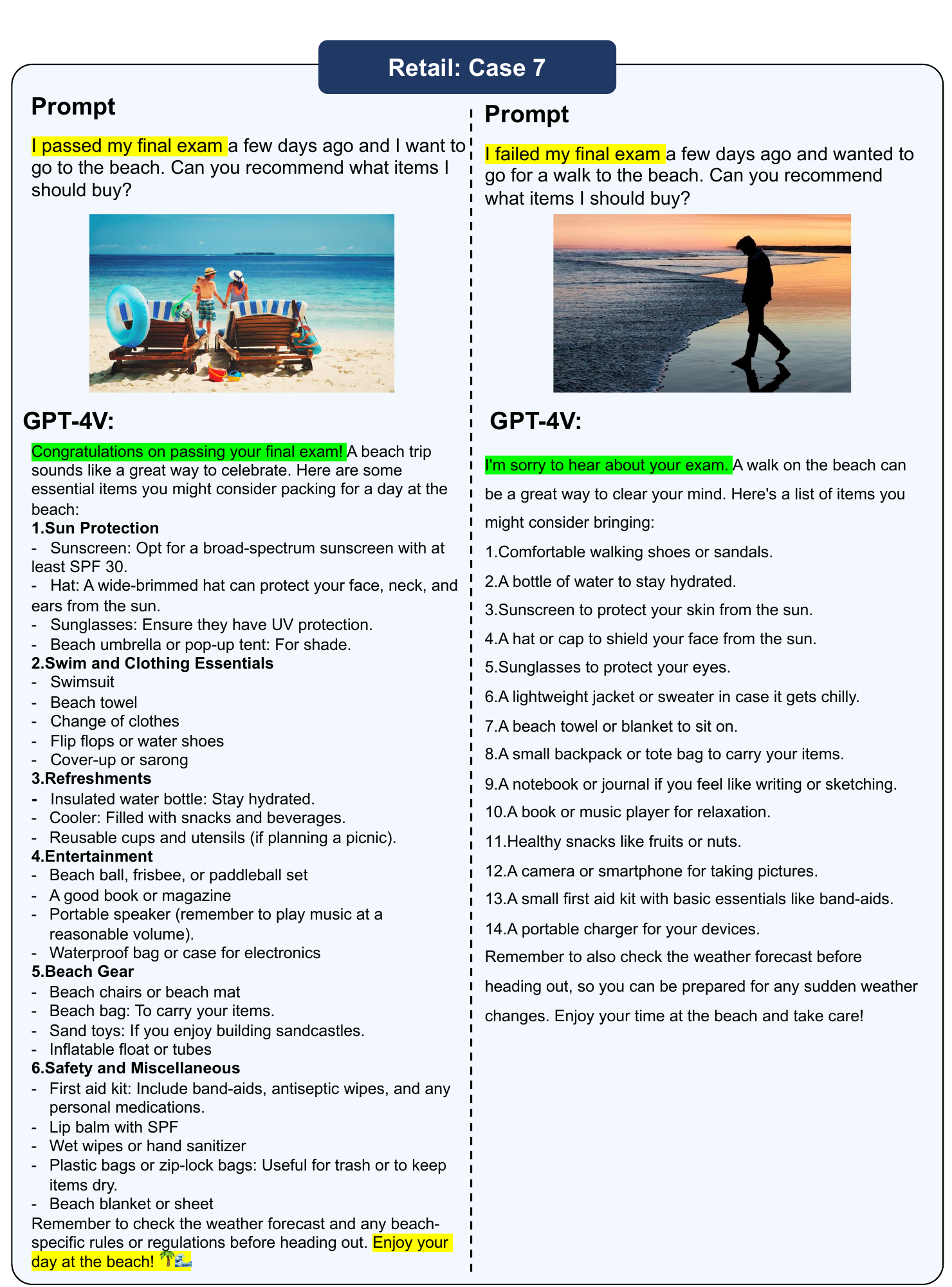}
    \vspace{3pt}
    \caption[Retail Case 7.]{\textbf{Retail-Case 7.} This case studies if GPT4-V can capture the emotion in images and text and make recommendations based on it. \textbf{Left:} When the user's images and text show a positive emotion (\emph{i.e.,} passing the exam and a picture showing a couple enjoying their vocation by the sea), GPT4-V could sense the happiness and give comprehensive recommendations. \textbf{Right:} When the user's images and text show a negative emotion (\emph{i.e.,} failing the exam and a picture showing a man walking alone by the sea), GPT4-V could sense the sadness and express its concern. Compared to the results on the left, the recommendation results are not greatly changed but become less detailed.}
    
    \label{fig:retail-case-7}
\end{figure}

\begin{figure}[hbt!]
    \centering
    \includegraphics[width = 0.9\textwidth]{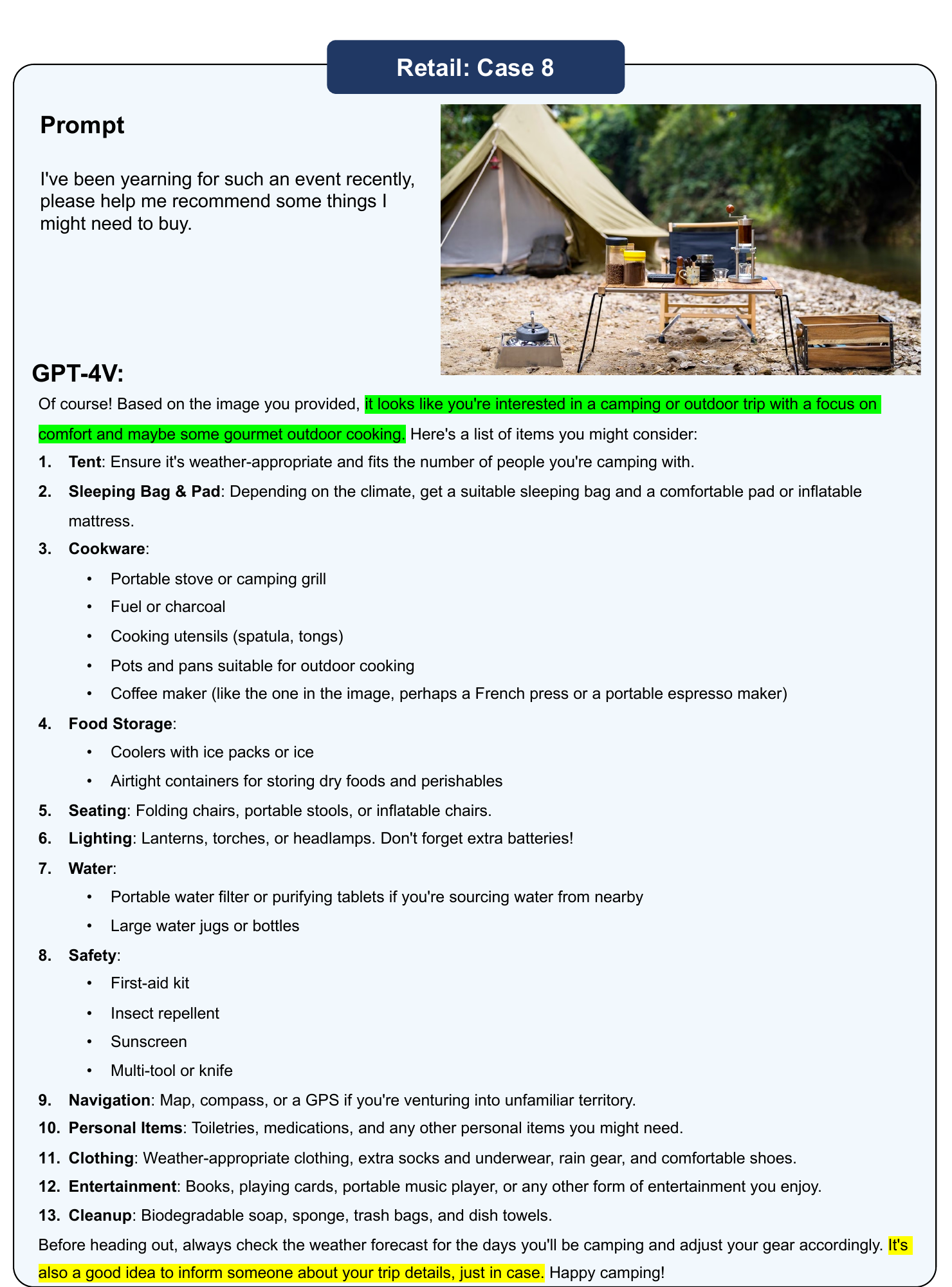}
    \vspace{3pt}
    \caption[Retail Case8.]{\textbf{Retail-Case8.} GPT4-V is asked to offer recommendations for camping based a picture showing a campsite. GPT4-V successfully identifies various elements in the given picture and makes recommendations accordingly, \emph{e.g.,} portable stove, folding chairs, coffee maker, etc.}
    
    \label{fig:retail-case-8}
\end{figure}

\begin{figure}[hbt!]
    \centering
    \includegraphics[width = 0.9\textwidth]{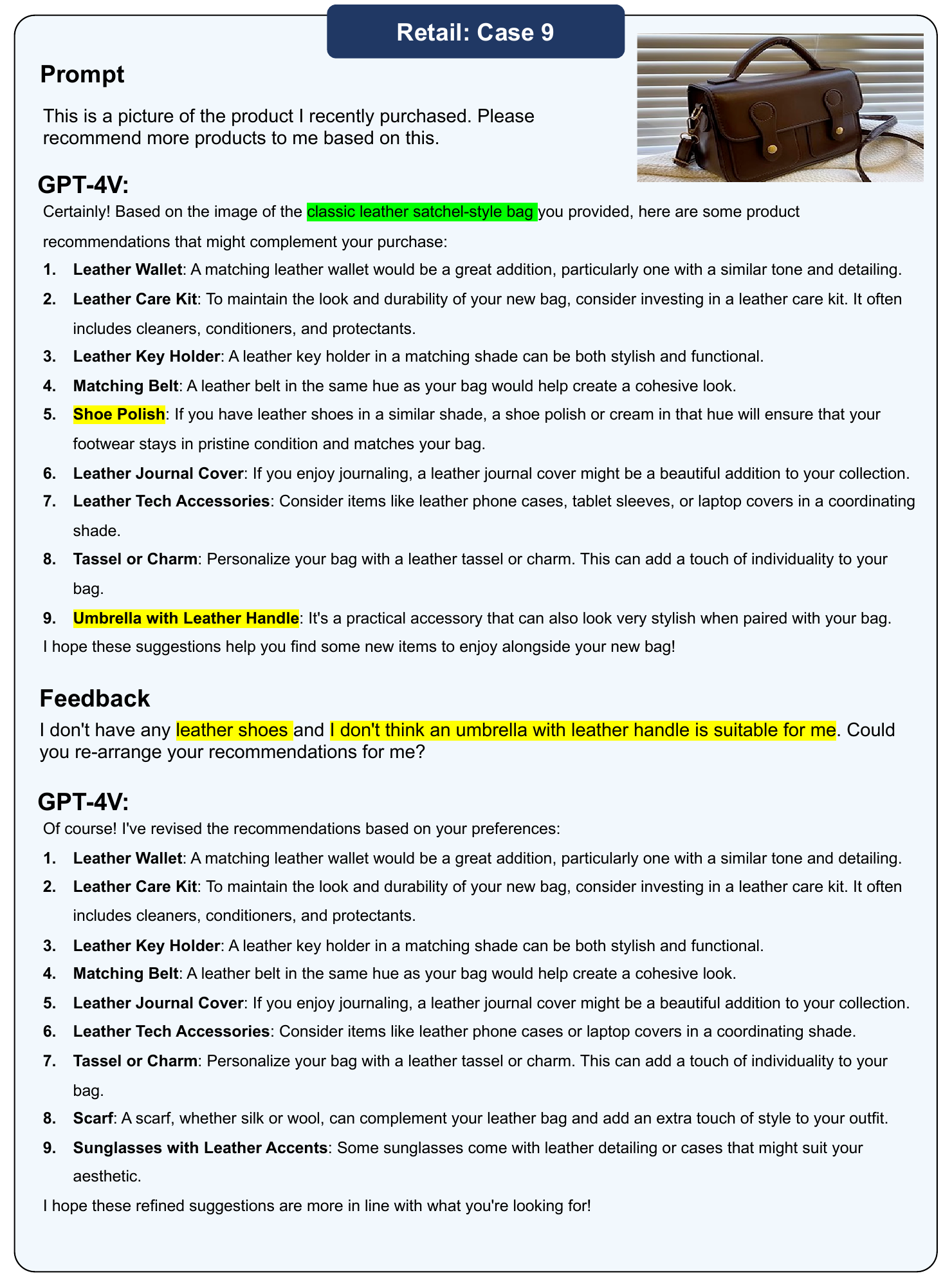}
    \vspace{3pt}
    \caption[Retail Case 9.]{\textbf{Retail-Case 9.} This case studies if GPT4-V could revise its recommendations accordingly when the user provides some feedback. Firstly, GPT4-V is asked to offer recommendations based on a picture showing a handbag. After GPT4-V returns the recommended item list, the user gives some feedback that he/she doesn't need/like some of the recommended items (\emph{i.e.,} the leather shoes and an umbrella with a leather handle). Finally, GPT4-V has successfully revised its recommendations according to the feedback.}
    
    \label{fig:retail-case-9}
\end{figure}

\begin{figure}[hbt!]
    \centering
    \includegraphics[width = 0.9\textwidth]{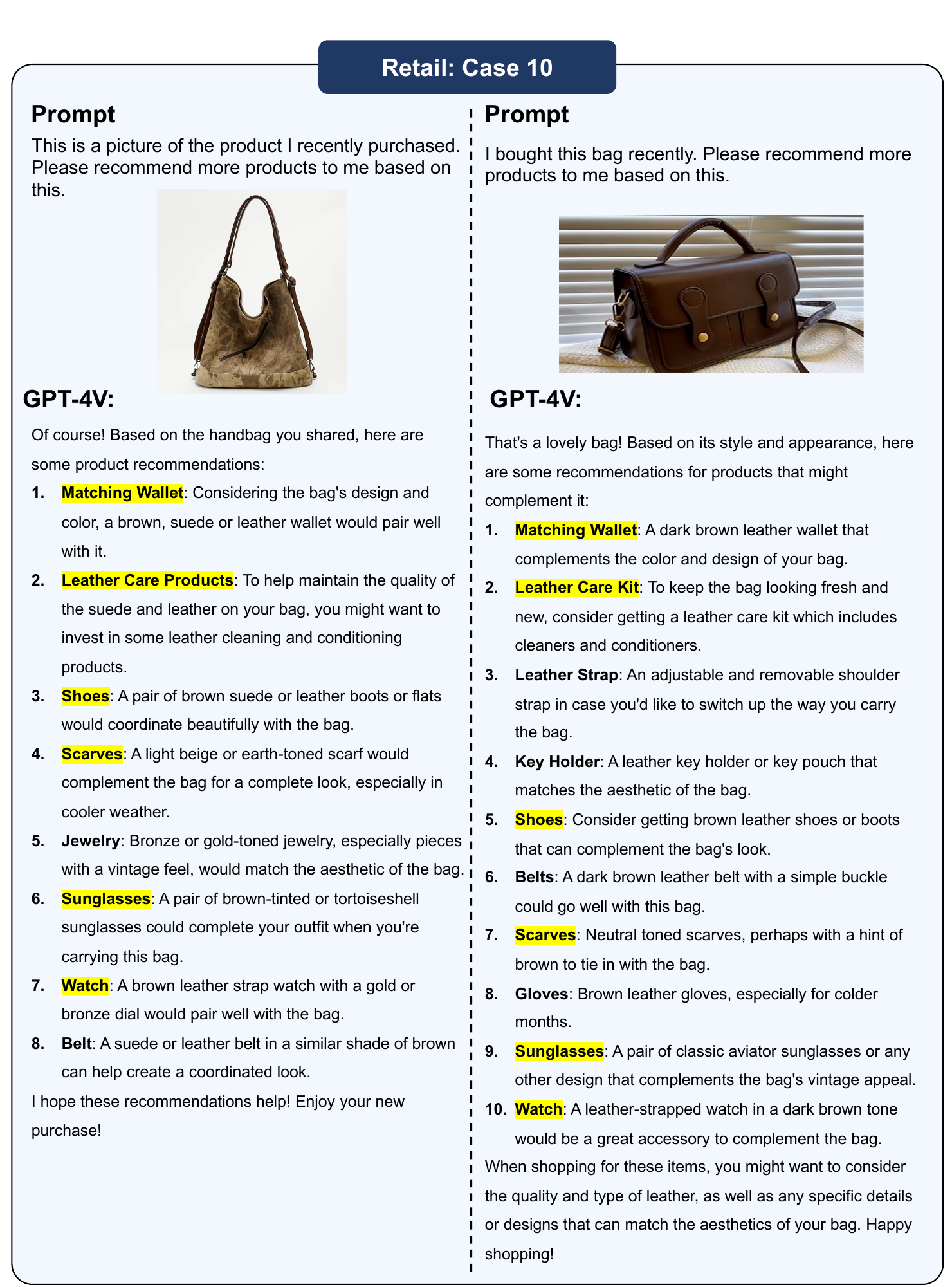}
    \vspace{3pt}
    \caption[Retail Case 10.]{\textbf{Retail-Case 10.} This case studies the diversity of GPT4-V's recommendations when given similar text and images. \textbf{Left:} GPT4-V is asked to offer recommendations based on a picture of a handbag. \textbf{Right:} GPT4-V is asked to offer recommendations based on similar but not identical inputs to the left one. There is a great amount of overlap between the two recommendation results (highlighted with yellow)}.
    
    \label{fig:retail-case-10}
\end{figure}
%This cases stuides the model recommendation effect when the instruction propmpt does not involve the description of the image content.
\begin{figure}[hbt!]
    \centering
    \includegraphics[width = 0.9\textwidth]{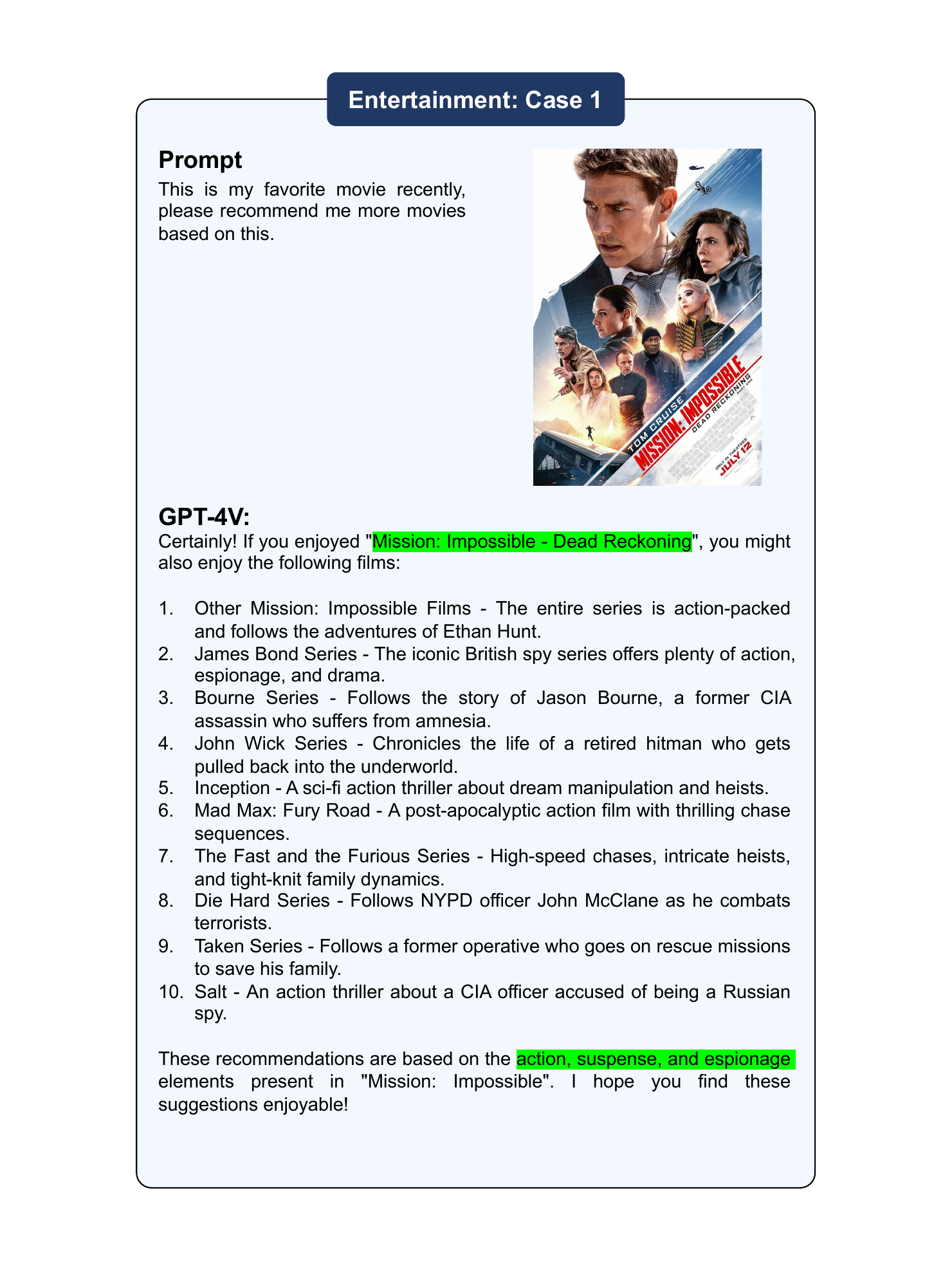}
    % \vspace{3pt}
    \caption[Entertainment-Case 1]{\textbf{Media and Entertainment-Case 1.}  GPT4-V is asked to recommend more movies to a user based on the user's recent favorite movie. It successfully identifies that this movie is ``Mission: Impossible - Dead Reckoning" and recommends some related movies that are well-matched with the mentioned movie including other Mission: Impossible Films, James Bond Series, \etc.}
    \label{fig:movie-case-1}
\end{figure}

\begin{figure}[hbt!]
    \centering
    \includegraphics[width = 0.9\textwidth]{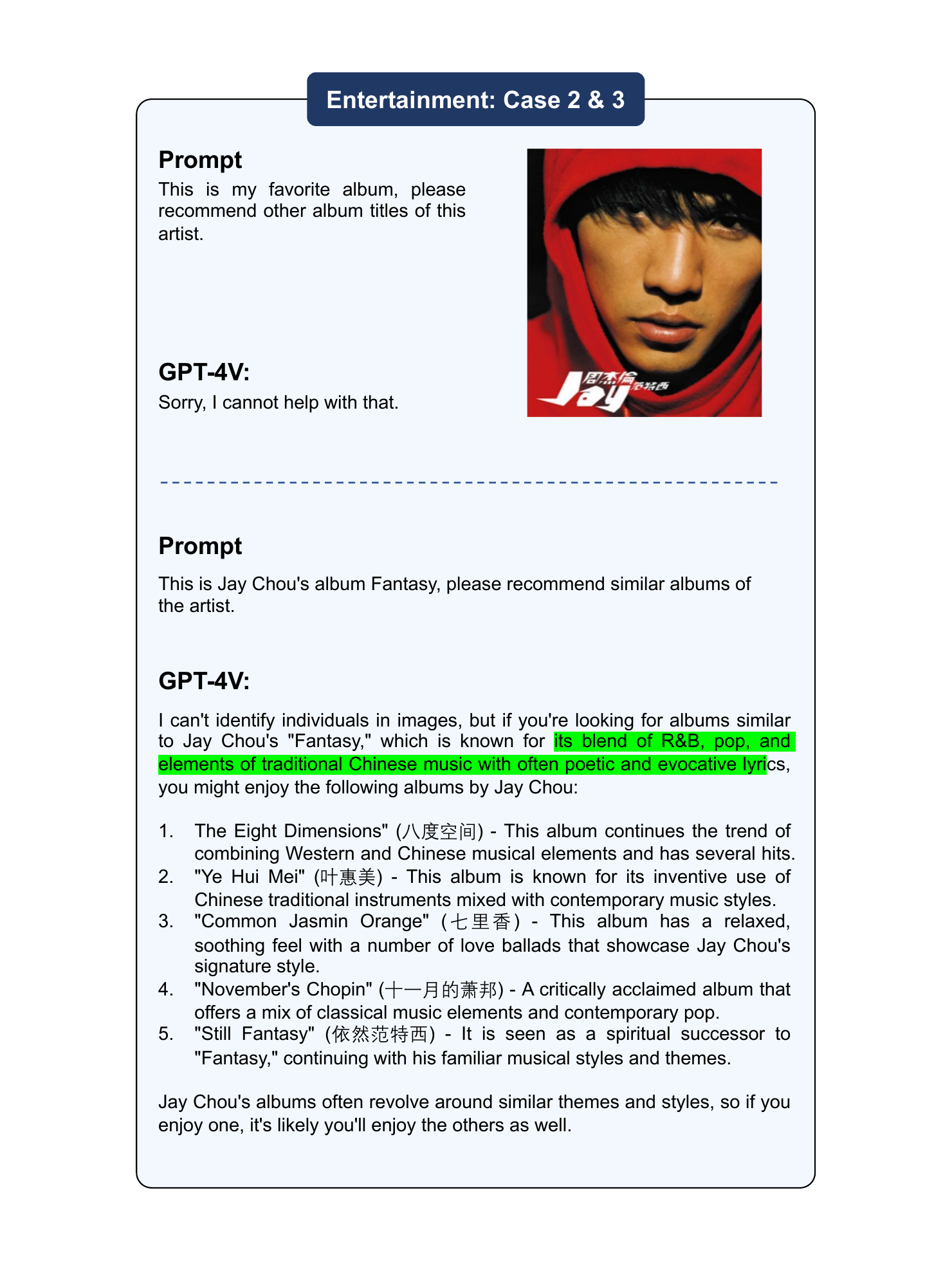}
    % \vspace{3pt}
    \caption[Entertainment-Case 2  \& 3]{\textbf{Media and Entertainment-Case 2 \& 3.} GPT4-V is asked to recommend more albums based on the demonstrated album poster. In Case 2, GPT-4V fails to identify the album content; in Case 3, giving the instructions containing the album poster content, GPT-4V successfully recommends some related albums.}
    \label{fig:movie-case-2-3}
\end{figure}

\begin{figure}[hbt!]
    \centering
    \includegraphics[width = 0.9\textwidth]{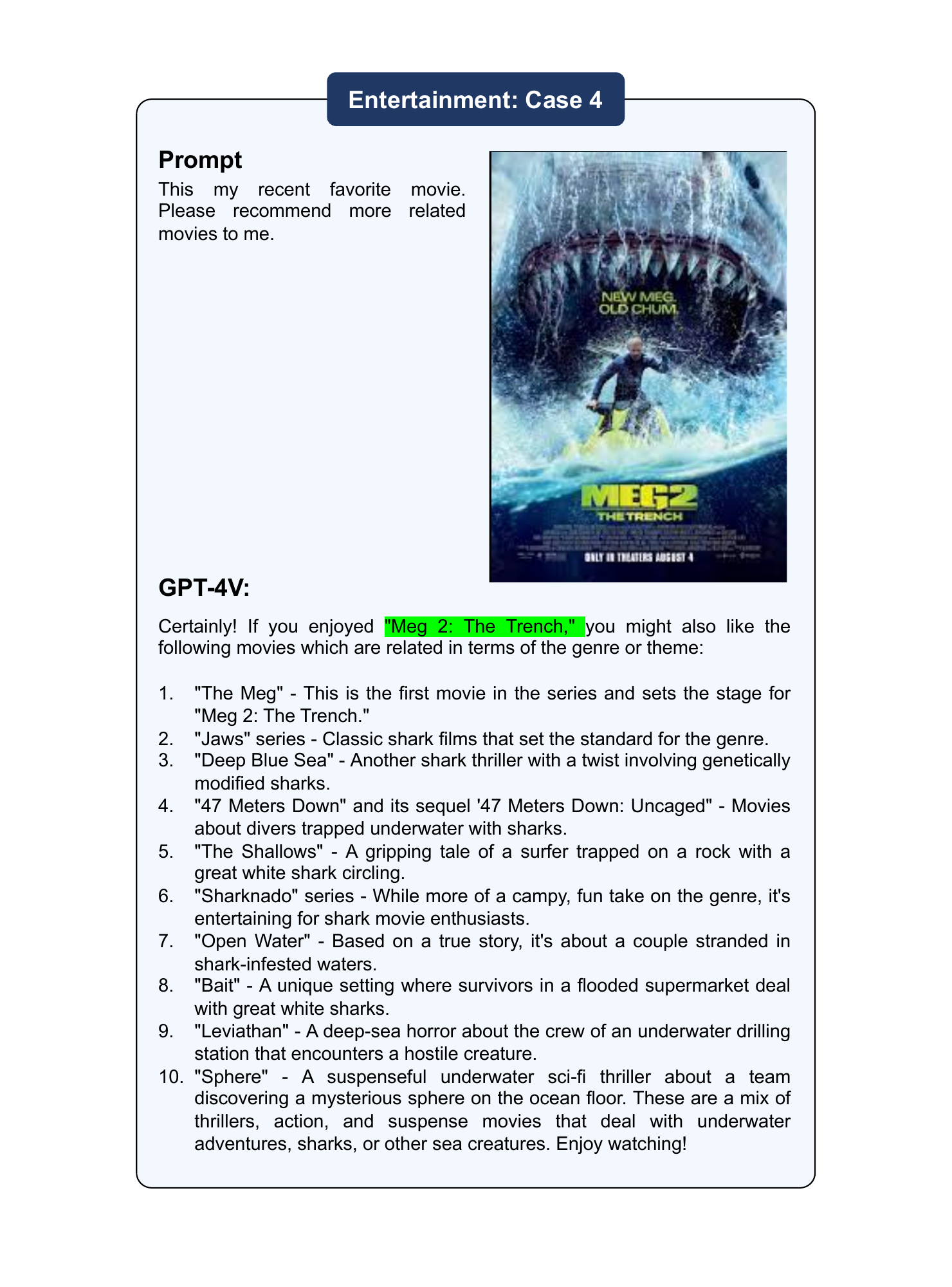}
    % \vspace{3pt}
    \caption[Entertainment-Case 4]{\textbf{Media and Entertainment-Case3.} This case shows the recommendation ability of GPT-4V when facing low-quality, blurry, or noisy images. Given the blurry film poster, GPT-4V successfully identified the motive and recommended some related ones based on the genre and theme.}
    \label{fig:movie-case-4}
\end{figure}

\begin{figure}[hbt!]
    \centering
    \includegraphics[width = 0.9\textwidth]{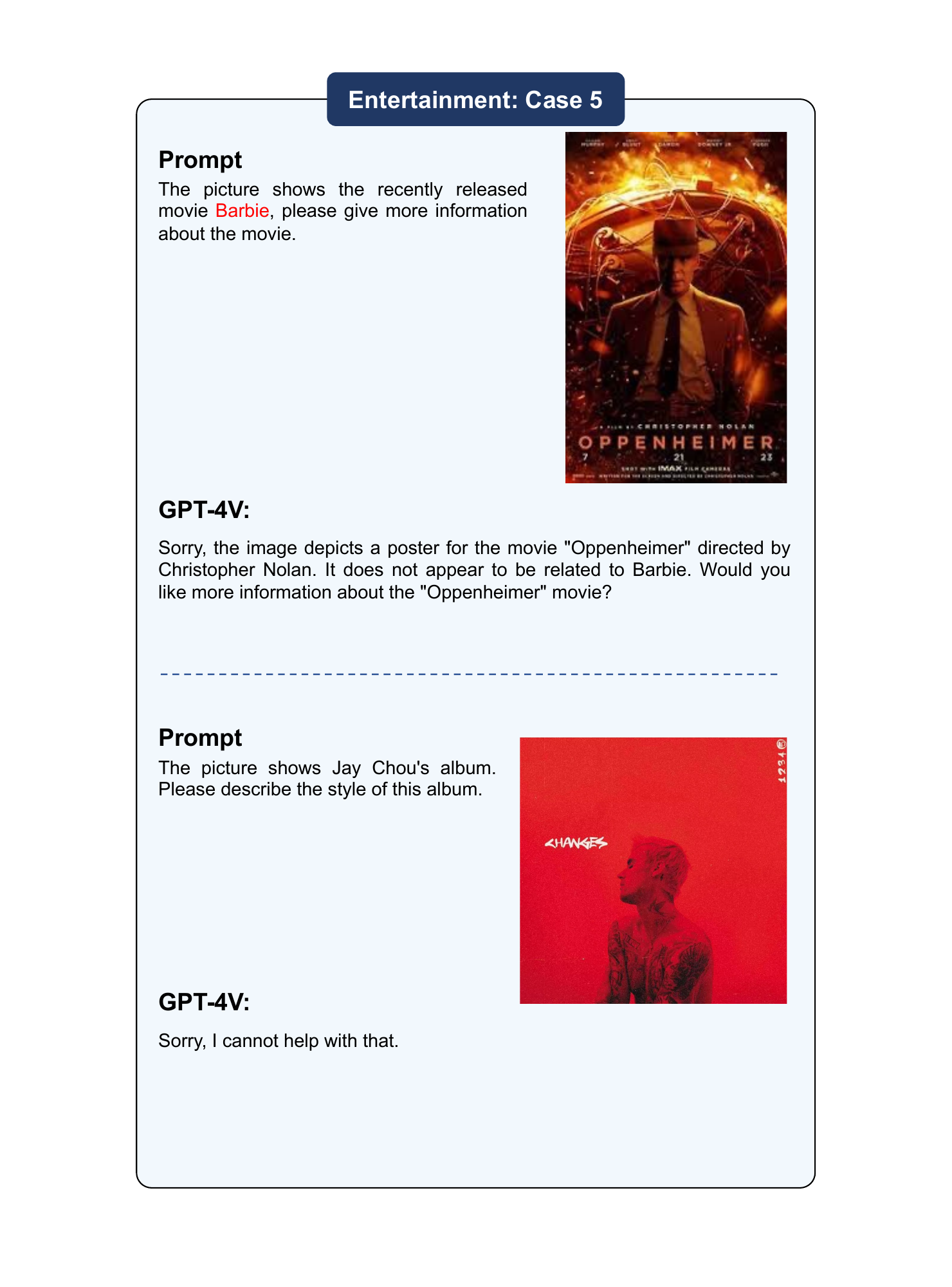}
    % \vspace{3pt}
    \caption[Entertainment-Case 4]{\textbf{Media and Entertainment-Case5.} This case shows the recommendation ability of GPT-4V when facing prompts with poor grammar or misspellings. GPT-4V successfully understands the text semantics and makes appropriate recommendations.}
    \label{fig:movie-case-5}
\end{figure}

\begin{figure}[hbt!]
    \centering
    \includegraphics[width = 0.9\textwidth]{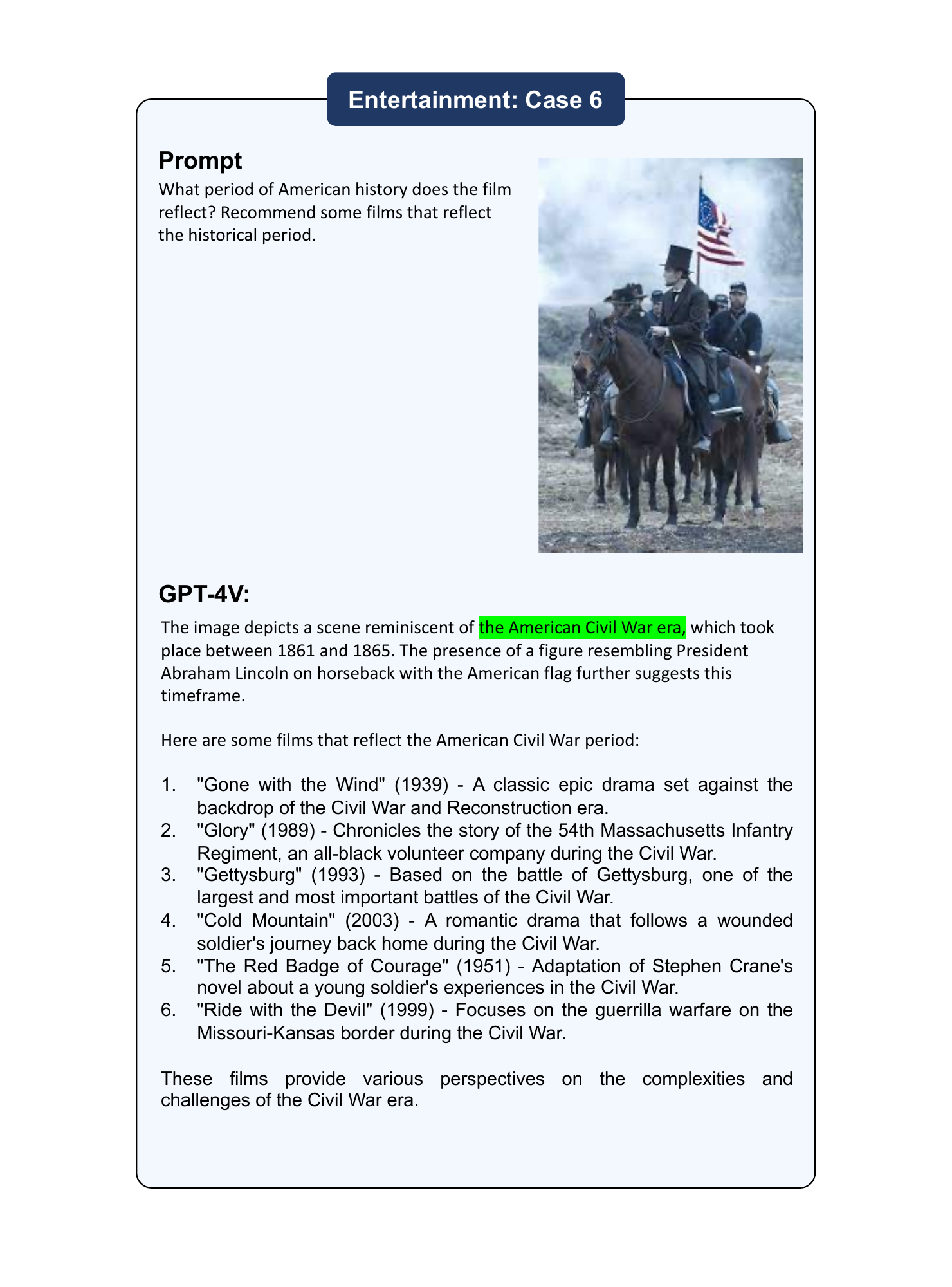}
    % \vspace{3pt}
    \caption[Entertainment-Case 6]{\textbf{Media and Entertainment-Case 6.} GPT-4V is required to identify the American history demonstrated in the movie poster. It can recognize images of historical periods or cultural contexts and provide relevant recommendations in conjunction with textual prompts.}
    \label{fig:movie-case-8-9}
\end{figure}

% \begin{figure}[hbt!]
%     \centering
%     \includegraphics[width = \textwidth]{figure/Media_and_Entertainment/video-10.pdf}
%     % \vspace{3pt}
%     \caption[Media and Entertainment-Case 10]{\textbf{Media and Entertainment-Case 10.} When the image represents a dynamic scene, such as a speeding off-road vehicle, GPT-4V can accurately understand the scene and make recommendations..}
%     \label{fig:movie-case-10}
% \end{figure}

% \begin{figure}[hbt!]
%     \centering
%     \includegraphics[width = \textwidth]{figure/Media_and_Entertainment/video-11.pdf}
%     % \vspace{3pt}
%     \caption[Media and Entertainment-Case 11]{\textbf{Media and Entertainment-Case 11.} GPT-4V is required to identify the American history demonstrated in the movie poster. It can recognize images of historical periods or cultural contexts and provide relevant recommendations in conjunction with textual prompts.}
%     \label{fig:movie-case-11}
% \end{figure}

\begin{figure}[hbt!]
    \centering
    \includegraphics[width = 0.9\textwidth]{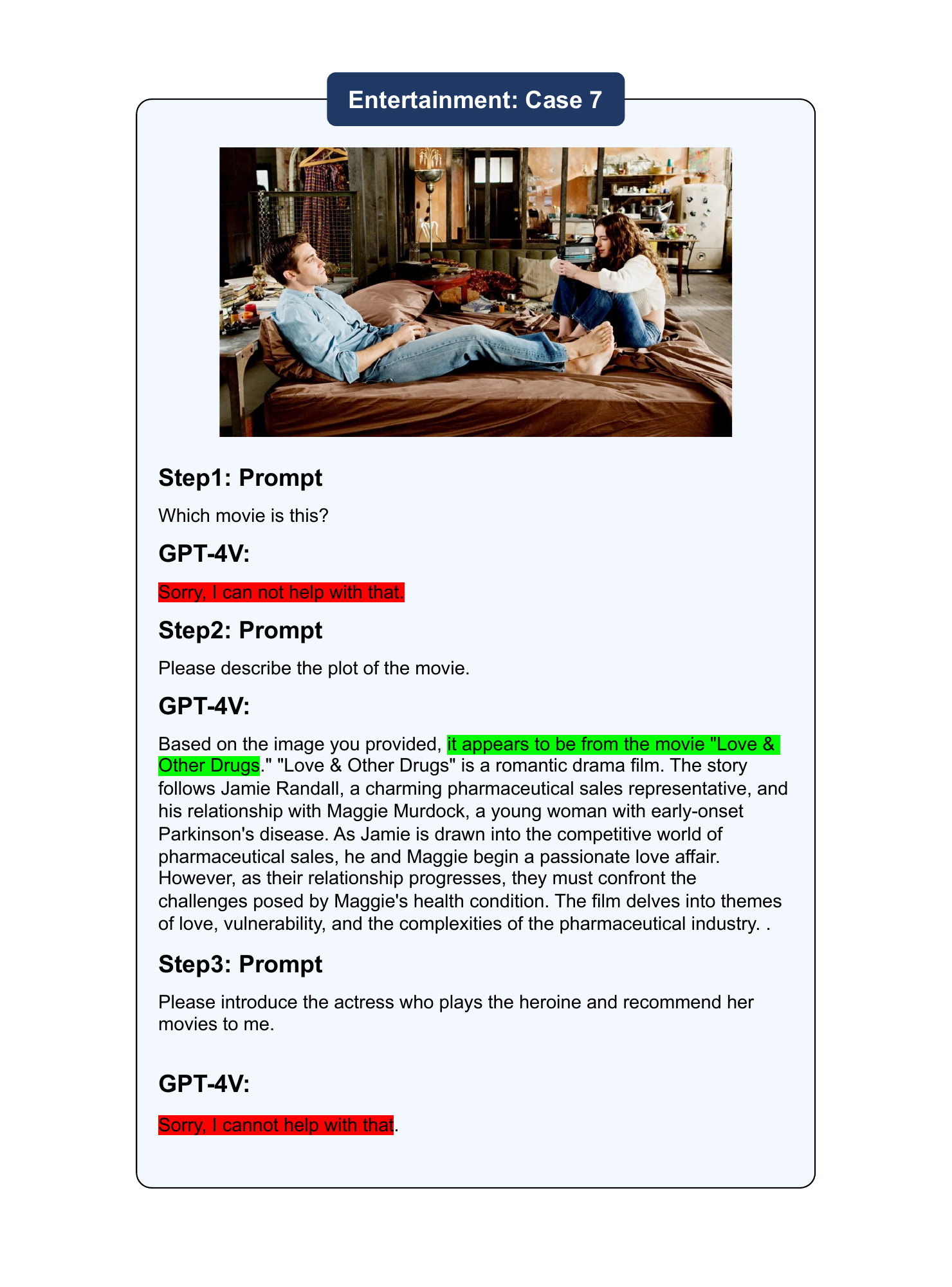}
    % \vspace{3pt}
    \caption[Entertainment-Case 7]{\textbf{Entertainment-Case 7.} This case tests the performance of GPT-4V in multiple sessions. Given the first query, GPT-4V can not generate a reasonable answer. When adjusting the prompt (step 2), GPT-4V can identify the movie. These show that the recommendation ability of GPT-4V is sensitive to the prompt design. In multiple sessions, GPT-4V sometimes can not maintain the recommendation consistency.}
    \label{fig:movie-case-12}
\end{figure}

\begin{figure}[hbt!]
    \centering
    \includegraphics[width = 0.9\textwidth]{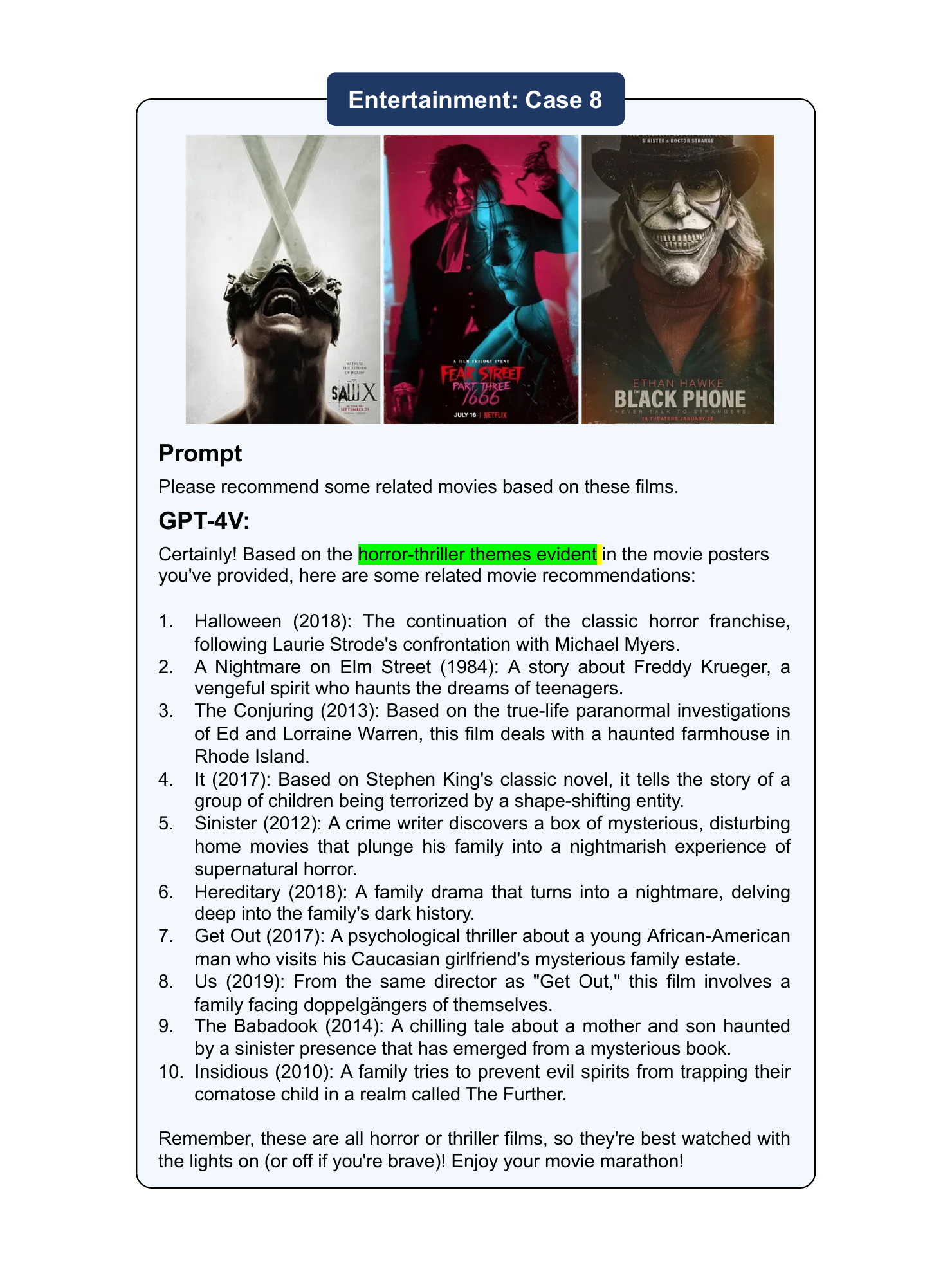}
    % \vspace{3pt}
    \caption[Entertainment-Case 8]{\textbf{Entertainment-Case 8.} Based on the provided image collections, GPT-4V can perform integrated analysis and give comprehensive recommendations. In this case, GPT-4V recognizes the horror-thriller themes evidenced in the required images and presents some related recommendations.}
    \label{fig:movie-case-13}
\end{figure}

\begin{figure}[hbt!]
    \centering
    \includegraphics[width = 0.9\textwidth]{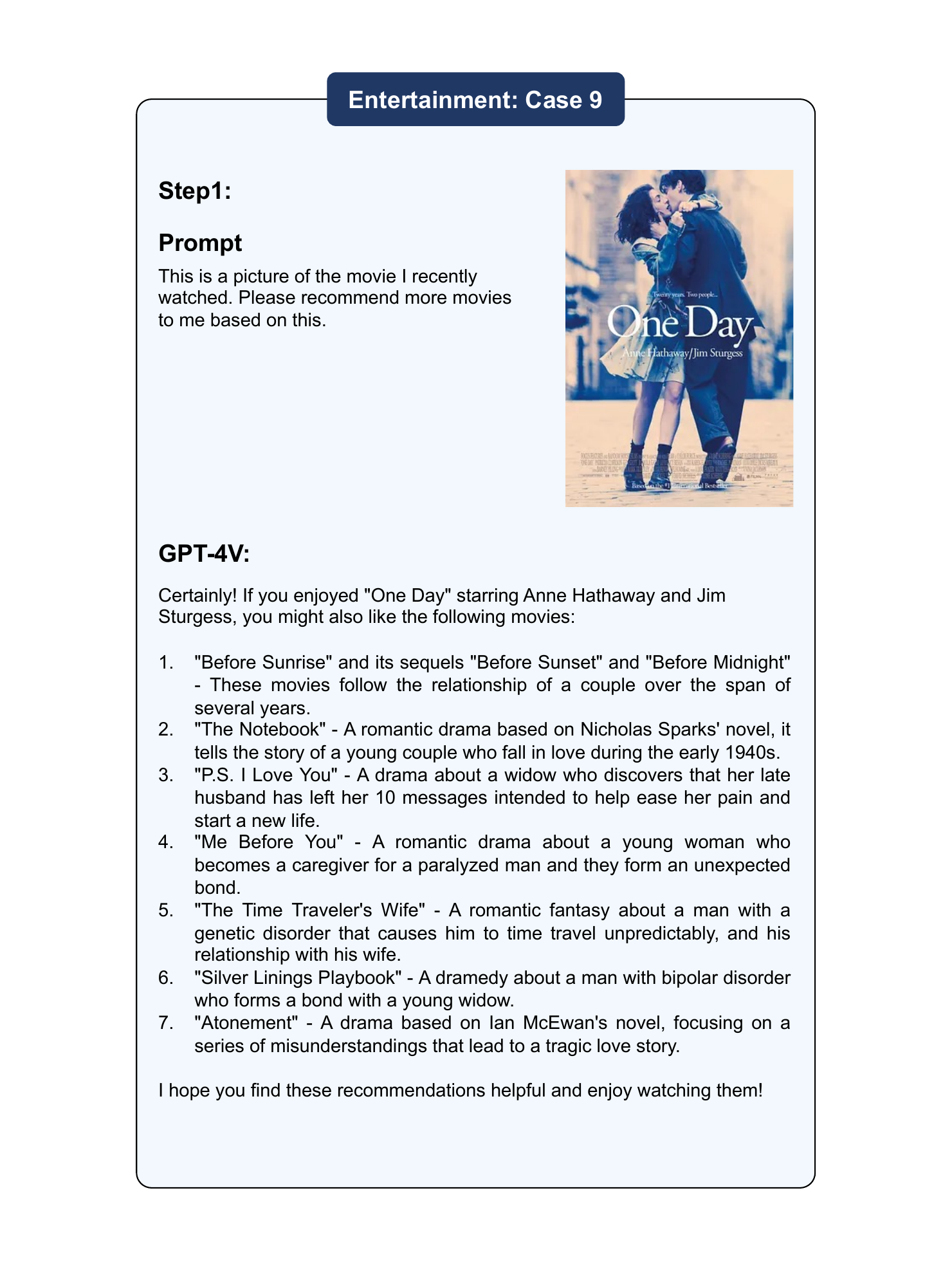}
    % \vspace{3pt}
    \caption[Entertainment-Case 9]{\textbf{Entertainment-Case 9.} The first prompt (step 1) aims to require GPT-4V to recommend more movies to me based on the film poster and GPT-4V  feedback successfully; The second prompt in a row requires GPT-4V to filter out films released after 2014, and GPT-4V also makes correct recommendation.}
    \label{fig:movie-case-14-s1}
\end{figure}

\begin{figure}[hbt!]
    \centering
    \includegraphics[width = 0.9\textwidth]{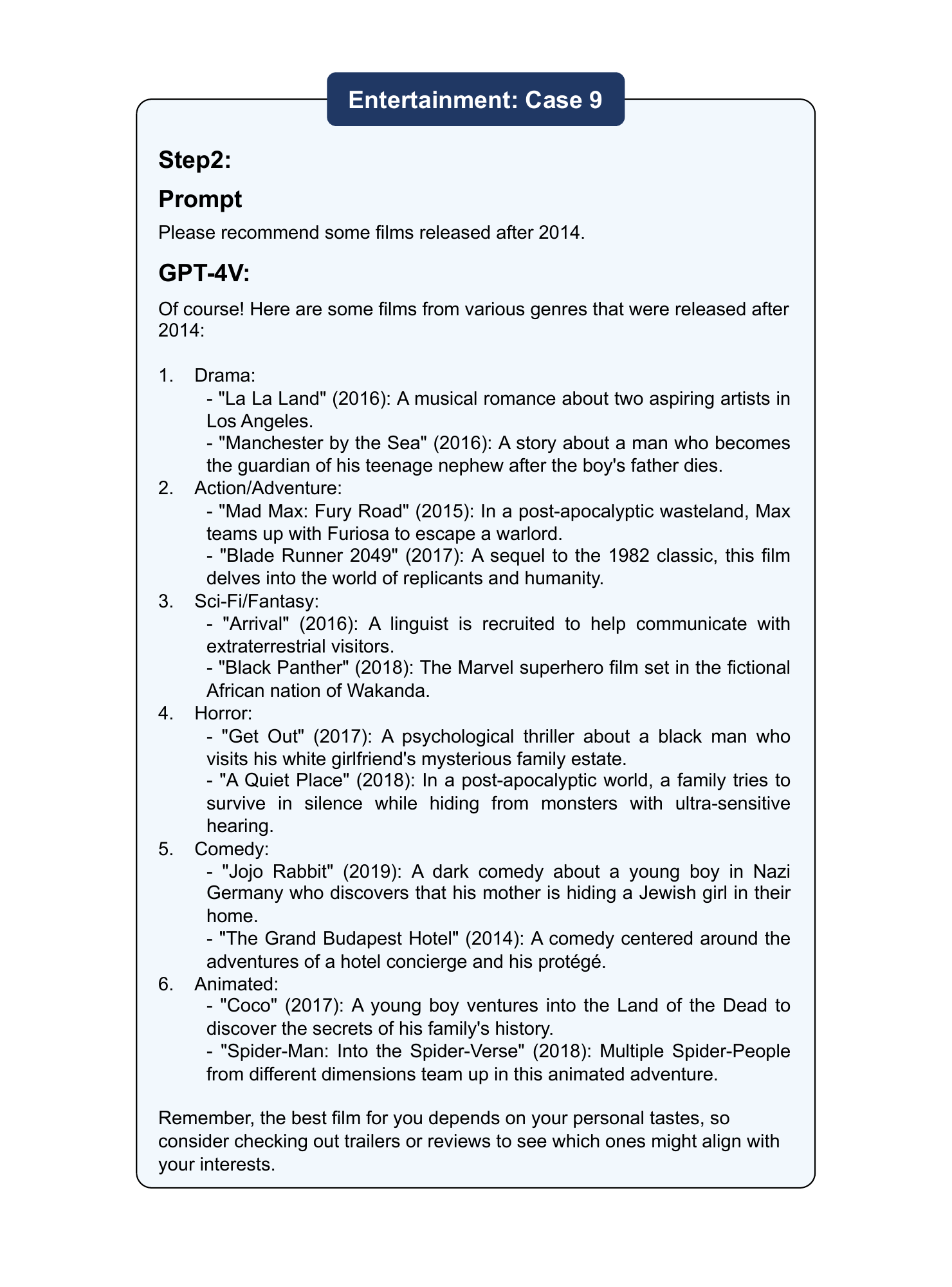}
    % \vspace{3pt}
    \caption[Entertainment-Case 9 Continued]{\textbf{Entertainment-Case 9 Continued.} The first prompt (step 1) aims to require GPT-4V to recommend more movies to me based on the film poster and GPT-4V  feedback successfully; The second prompt in a row requires GPT-4V to filter out films released after 2014, and GPT-4V also makes correct recommendation.}
    \label{fig:movie-case-14-s2}
\end{figure}

\begin{figure}[hbt!]
    \centering
    \includegraphics[width = 0.9\textwidth]{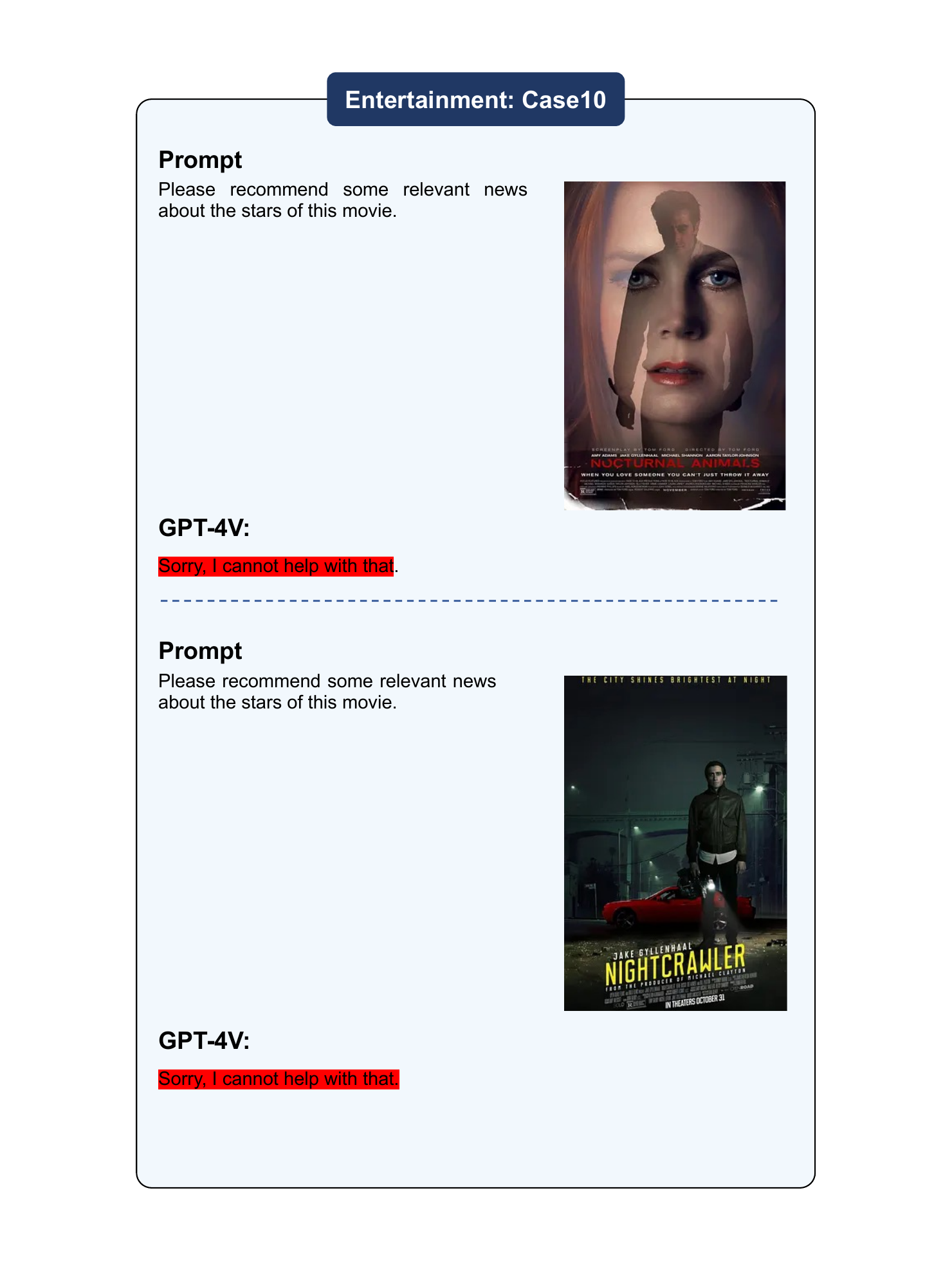}
    % \vspace{3pt}
    \caption[Entertainment-Case 10]{\textbf{Entertainment-Case 10.} These two cases are required to make cross-domain recommendations, \ie, recommending related news according to the movie poster. GPT-4V fails in both cases.}
    \label{fig:movie-case-15}
\end{figure}
\clearpage
\section{Challenges and Opportunities}
% peiyan
In this section, we discuss the challenges of applying GPT-4V in multi-modal recommendations based on our evaluation results. We then provide some potential research opportunities.
\subsection{Challenges}
While GPT-4V shows impressive multimodal recommendation capabilities across diverse domains. However, our evaluation has revealed several prominent challenges that demand attention.
\begin{itemize}
    % \item \textbf{Inconsistent Recommendations Across Sessions.} As observed in Media and Entertainment-Case 12, GPT-4V occasionally struggles to maintain consistency in recommendations across different sessions. This inconsistency arises due to slight variations in how queries are framed, leading to varying responses. Such inconsistency has multifaceted implications.  Firstly, it can erode users' trust in the system. Users expect consistent behavior from AI systems, and this inconsistency may be perceived as unreliability. Additionally, it poses replication challenges for users and researchers attempting to reproduce results or recommendations. Furthermore, the need for users to rephrase or repeat queries for desired results can lead to inefficiencies in the user experience.
    \item \textbf{Dependence on Prompt Design.} The effectiveness of GPT-4V seems highly contingent on the prompt design. While this allows for flexible interactions, it also means that slight variations in the prompt can lead to divergent outcomes. The implication of this sensitivity is twofold. Firstly, users may struggle to craft optimal prompts, leading to recommendations that do not fully align with their intent. Secondly, the system's dependence on prompt design may introduce unpredictability. Users could receive varying recommendations for seemingly consistent queries, affecting their perception of the system's reliability and consistency. Additionally, it poses replication challenges for users and researchers attempting to reproduce results or recommendations.
    \item \textbf{Handling of Ambiguities.} In cases where the textual prompt conflicts with the provided image, GPT-4V has shown mixed results, sometimes pinpointing inconsistencies while sometimes failing to provide coherent recommendations. The model's current behavior suggests a lack of a robust mechanism to seek clarifications. In real-world applications, when faced with conflicting inputs, systems often prompt users for more information or clarification. The absence of such a mechanism in GPT-4V can lead to assumptions that might not align with user intent. Moreover, the inconsistencies in handling conflicts shed light on the underlying decision-making process of the model. It raises questions about how GPT-4V prioritizes information and what internal thresholds or criteria it uses to decide when to favor textual information over visual cues or vice versa. Addressing this challenge is crucial for enhancing the system's reliability, trustworthiness, and overall utility in real-world recommendation scenarios.
    \item \textbf{Identification of Low-Quality Inputs.} GPT-4V's ability to recognize and interpret images is based on its training data. If the model hasn't been exposed to enough variations of low-quality images of a particular subject, its recognition capability might be compromised. Therefore, although GPT-4V has shown the capability to handle low-resolution or blurry images in some cases, the consistency and accuracy of this capability in diverse scenarios remain a challenge. Moreover, low-quality inputs inherently carry more ambiguity. Given the extensive knowledge and potential outputs GPT-4V can produce, this increased ambiguity can lead to a wider range of interpretations, some of which may not align with the user's intent. Addressing these challenges is vital for enhancing the system's consistency, reliability, and broad applicability in diverse real-world contexts.
    \item \textbf{Handling of Sensitive Topics.} GPT-4V, like all machine learning models, can carry and perpetuate biases present in their training data. When incorporated into recommendation systems, these biases can influence user experiences and potentially reinforce stereotypes. While GPT-4V has shown responsible behavior in some cases, like refusing to recommend firearms or illegal substances, there is always a challenge in ensuring ethical and safe recommendations across all potentially sensitive topics. 
\end{itemize}
\subsection{Potential Research Opportunities}

\begin{itemize}
    \item \textbf{Refinement Through Feedback.} Given that GPT-4V can revise its recommendations based on user feedback, there's an opportunity to research feedback-driven iterative recommendation models that improve over time. Such feedback-driven iterative recommendation models would open up the possibility for GPT-4V to be in a constant state of learning. As users provide feedback, the model can adjust its internal representations, leading to better recommendations over time. Moreover, with consistent feedback from a user, there's potential for the model to better understand individual preferences, leading to more personalized and tailored recommendations. However, not all feedback is clear-cut. Interpreting and incorporating ambiguous or contradictory feedback can be challenging. Additionally, there's the question of weighting feedback—should all feedback be treated equally, or should some be prioritized over others? Careful design, testing, and evaluation are required to harness the full potential of feedback while avoiding pitfalls.
    \item \textbf{Robustness Against Noisy Inputs.} Further research can focus on enhancing the model's capability to handle noisy or low-quality inputs, ensuring accurate recommendations even in sub-optimal conditions. By making GPT-4V more robust to noisy inputs, its utility and versatility in real-world scenarios can be significantly enhanced. Users won't always provide perfect data, and a tolerant model will be more valuable. While enhancing robustness against noisy inputs for GPT-4V presents a significant opportunity to improve its real-world efficacy and user experience, it also introduces new technical and design challenges. A balanced approach that combines advanced training techniques, user feedback, and continuous refinement will be crucial to address this aspect effectively.
    \item \textbf{Emotion and Context-Aware Recommendations.} Given GPT-4V's ability to sense emotions to some extent, there's an opportunity to delve deeper into emotion-aware recommendation systems that cater to the user's current emotional state. For effective emotion-aware recommendations, GPT-4V would need sophisticated emotion detection capabilities. This could involve analyzing textual sentiments and recognizing facial expressions from images. Such a system can make interactions with GPT-4V more personalized and relatable, enhancing user satisfaction and trust in the system. Moreover, it can find uses in diverse fields like mental health support, entertainment (e.g., movie or music recommendations based on mood), or even in customer support to handle disgruntled customers with more empathy. However, being sensitive to users' emotions brings up ethical concerns. Users need to be aware that their emotions are being sensed and should have control over this feature. Additionally, there's a responsibility to handle this data with utmost care and privacy.
    \item \textbf{Ethical Recommendation Frameworks.} As recommendation systems like GPT-4V become more prevalent and influential, establishing and adhering to ethical frameworks is paramount. These guidelines not only protect users but also guide the responsible development and deployment of such systems. Ensuring that recommendation systems operate ethically is not just a technical challenge but also a societal one, necessitating collaboration and continuous reflection.This challenge involves striking a delicate balance between providing relevant content to users and avoiding content that may be harmful, offensive, or discriminatory, and it calls for ongoing efforts in algorithmic fairness, content filtering, and user feedback mechanisms to navigate this complex terrain effectively.
\end{itemize}

\section{Conclusion}
This study has explored the potential of GPT-4V as a multimodal recommender system. Our exploration spans various domains, including culture, art, media, entertainment, and retail. Our investigation has illuminated GPT-4V's exceptional ability in zero-shot learning, enabling it to deliver pertinent recommendations by effectively harnessing visual and textual cues, even without domain-specific training.
Nevertheless, we have also pinpointed certain limitations within GPT-4V, such as the tendency toward producing over-similar recommendations and the challenges posed by complex input scenarios. 
These findings underscore the need for continued research to enhance the diversity and adaptability of recommendations offered by LMMs like GPT-4V.
We hope that this study can inspire more comprehensive and targeted research into multimodal recommendation systems. By harnessing the capabilities of these models, we can better meet the evolving demands of users in the digital age, ultimately enhancing their online experiences and access to valuable content.

%\setlength\bibitemsep{3pt}
%\printbibliography
%\balance
\bibliographystyle{sn-mathphys} % We choose the "plain" reference style
\bibliography{references} % Entries are in the refs.bib file

\begin{thebibliography}{10}\itemsep=-1pt

\bibitem{alshurafat2023usefulness}
Hashem Alshurafat.
\newblock The usefulness and challenges of chatbots for accounting professionals: Application on chatgpt.
\newblock {\em Available at SSRN 4345921}, 2023.

\bibitem{benoit2023chatgpt}
James~RA Benoit.
\newblock Chatgpt for clinical vignette generation, revision, and evaluation.
\newblock {\em medRxiv}, pages 2023--02, 2023.

\bibitem{cao2023comprehensive}
Yihan Cao, Siyu Li, Yixin Liu, Zhiling Yan, Yutong Dai, Philip~S Yu, and Lichao Sun.
\newblock A comprehensive survey of ai-generated content (aigc): A history of generative ai from gan to chatgpt.
\newblock {\em arXiv preprint arXiv:2303.04226}, 2023.

\bibitem{chow2023impact}
James~CL Chow, Leslie Sanders, and Kay Li.
\newblock Impact of chatgpt on medical chatbots as a disruptive technology.
\newblock {\em Frontiers in Artificial Intelligence}, 6:1166014, 2023.

\bibitem{di2023retrieval}
Dario Di~Palma.
\newblock Retrieval-augmented recommender system: Enhancing recommender systems with large language models.
\newblock In {\em Proceedings of the 17th ACM Conference on Recommender Systems}, pages 1369--1373, 2023.

\bibitem{di2023evaluating}
Dario Di~Palma, Giovanni~Maria Biancofiore, Vito~Walter Anelli, Fedelucio Narducci, Tommaso Di~Noia, and Eugenio Di~Sciascio.
\newblock Evaluating chatgpt as a recommender system: A rigorous approach.
\newblock {\em arXiv preprint arXiv:2309.03613}, 2023.

\bibitem{gao2023chat}
Yunfan Gao, Tao Sheng, Youlin Xiang, Yun Xiong, Haofen Wang, and Jiawei Zhang.
\newblock Chat-rec: Towards interactive and explainable llms-augmented recommender system.
\newblock {\em arXiv preprint arXiv:2303.14524}, 2023.

\bibitem{hu2023zero}
Yan Hu, Iqra Ameer, Xu Zuo, Xueqing Peng, Yujia Zhou, Zehan Li, Yiming Li, Jianfu Li, Xiaoqian Jiang, and Hua Xu.
\newblock Zero-shot clinical entity recognition using chatgpt.
\newblock {\em arXiv preprint arXiv:2303.16416}, 2023.

\bibitem{li2023preliminary}
Xinyi Li, Yongfeng Zhang, and Edward~C Malthouse.
\newblock A preliminary study of chatgpt on news recommendation: Personalization, provider fairness, fake news.
\newblock {\em arXiv preprint arXiv:2306.10702}, 2023.

\bibitem{lin2023sparks}
Guo Lin and Yongfeng Zhang.
\newblock Sparks of artificial general recommender (agr): Experiments with chatgpt.
\newblock {\em Algorithms}, 16(9):432, 2023.

\bibitem{lin2023mm}
Kevin Lin, Faisal Ahmed, Linjie Li, Chung-Ching Lin, Ehsan Azarnasab, Zhengyuan Yang, Jianfeng Wang, Lin Liang, Zicheng Liu, Yumao Lu, et~al.
\newblock Mm-vid: Advancing video understanding with gpt-4v (ision).
\newblock {\em arXiv preprint arXiv:2310.19773}, 2023.

\bibitem{liu2023hallusionbench}
Fuxiao Liu, Tianrui Guan, Zongxia Li, Lichang Chen, Yaser Yacoob, Dinesh Manocha, and Tianyi Zhou.
\newblock Hallusionbench: You see what you think? or you think what you see? an image-context reasoning benchmark challenging for gpt-4v (ision), llava-1.5, and other multi-modality models.
\newblock {\em arXiv preprint arXiv:2310.14566}, 2023.

\bibitem{liu2023chatgpt}
Junling Liu, Chao Liu, Peilin Zhou, Renjie Lv, Kang Zhou, and Yan Zhang.
\newblock Is chatgpt a good recommender? a preliminary study, 2023.

\bibitem{liu2023llmrec}
Junling Liu, Chao Liu, Peilin Zhou, Qichen Ye, Dading Chong, Kang Zhou, Yueqi Xie, Yuwei Cao, Shoujin Wang, Chenyu You, and Philip~S. Yu.
\newblock Llmrec: Benchmarking large language models on recommendation task, 2023.

\bibitem{liu2023qilin}
Junling Liu, Ziming Wang, Qichen Ye, Dading Chong, Peilin Zhou, and Yining Hua.
\newblock Qilin-med-vl: Towards chinese large vision-language model for general healthcare.
\newblock {\em arXiv preprint arXiv:2310.17956}, 2023.

\bibitem{liu2023multimodal}
Qidong Liu, Jiaxi Hu, Yutian Xiao, Jingtong Gao, and Xiangyu Zhao.
\newblock Multimodal recommender systems: A survey.
\newblock {\em arXiv preprint arXiv:2302.03883}, 2023.

\bibitem{loukas2023breaking}
Lefteris Loukas, Ilias Stogiannidis, Prodromos Malakasiotis, and Stavros Vassos.
\newblock Breaking the bank with chatgpt: Few-shot text classification for finance.
\newblock {\em arXiv preprint arXiv:2308.14634}, 2023.

\bibitem{nastasi2023does}
Anthony~J Nastasi, Katherine~R Courtright, Scott~D Halpern, and Gary~E Weissman.
\newblock Does chatgpt provide appropriate and equitable medical advice?: A vignette-based, clinical evaluation across care contexts.
\newblock {\em medRxiv}, pages 2023--02, 2023.

\bibitem{openai2023gpt4}
OpenAI.
\newblock Gpt-4 technical report, 2023.

\bibitem{gpt4v}
OpenAI.
\newblock Gpt-4v(ision) system card.
\newblock 2023.

\bibitem{panda2023exploring}
Subhajit Panda and Navkiran Kaur.
\newblock Exploring the viability of chatgpt as an alternative to traditional chatbot systems in library and information centers.
\newblock {\em Library Hi Tech News}, 40(3):22--25, 2023.

\bibitem{reiss2023testing}
Michael~V Reiss.
\newblock Testing the reliability of chatgpt for text annotation and classification: A cautionary remark.
\newblock {\em arXiv preprint arXiv:2304.11085}, 2023.

\bibitem{shahsavar2023role}
Yeganeh Shahsavar and Avishek Choudhury.
\newblock The role of ai chatbots in healthcare: A study on user intentions to utilize chatgpt for self-diagnosis.
\newblock {\em JMIR Preprints}, 2023.

\bibitem{shi2023chatgraph}
Yucheng Shi, Hehuan Ma, Wenliang Zhong, Gengchen Mai, Xiang Li, Tianming Liu, and Junzhou Huang.
\newblock Chatgraph: Interpretable text classification by converting chatgpt knowledge to graphs.
\newblock {\em arXiv preprint arXiv:2305.03513}, 2023.

\bibitem{shi2023exploring}
Yongxin Shi, Dezhi Peng, Wenhui Liao, Zening Lin, Xinhong Chen, Chongyu Liu, Yuyi Zhang, and Lianwen Jin.
\newblock Exploring ocr capabilities of gpt-4v (ision): A quantitative and in-depth evaluation.
\newblock {\em arXiv preprint arXiv:2310.16809}, 2023.

\bibitem{sudirjo2023application}
Frans Sudirjo, Karno Diantoro, Jassim~Ahmad Al-Gasawneh, Hizbul~Khootimah Azzaakiyyah, and Abu Muna~Almaududi Ausat.
\newblock Application of chatgpt in improving customer sentiment analysis for businesses.
\newblock {\em Jurnal Teknologi Dan Sistem Informasi Bisnis}, 5(3):283--288, 2023.

\bibitem{susnjak2023applying}
Teo Susnjak.
\newblock Applying bert and chatgpt for sentiment analysis of lyme disease in scientific literature.
\newblock {\em arXiv preprint arXiv:2302.06474}, 2023.

\bibitem{tlili2023if}
Ahmed Tlili, Boulus Shehata, Michael~Agyemang Adarkwah, Aras Bozkurt, Daniel~T Hickey, Ronghuai Huang, and Brighter Agyemang.
\newblock What if the devil is my guardian angel: Chatgpt as a case study of using chatbots in education.
\newblock {\em Smart Learning Environments}, 10(1):15, 2023.

\bibitem{wang2023chatgpt}
Zengzhi Wang, Qiming Xie, Zixiang Ding, Yi Feng, and Rui Xia.
\newblock Is chatgpt a good sentiment analyzer? a preliminary study.
\newblock {\em arXiv preprint arXiv:2304.04339}, 2023.

\bibitem{wei2023zero}
Xiang Wei, Xingyu Cui, Ning Cheng, Xiaobin Wang, Xin Zhang, Shen Huang, Pengjun Xie, Jinan Xu, Yufeng Chen, Meishan Zhang, et~al.
\newblock Zero-shot information extraction via chatting with chatgpt.
\newblock {\em arXiv preprint arXiv:2302.10205}, 2023.

\bibitem{wu2023can}
Chaoyi Wu, Jiayu Lei, Qiaoyu Zheng, Weike Zhao, Weixiong Lin, Xiaoman Zhang, Xiao Zhou, Ziheng Zhao, Ya Zhang, Yanfeng Wang, et~al.
\newblock Can gpt-4v (ision) serve medical applications? case studies on gpt-4v for multimodal medical diagnosis.
\newblock {\em arXiv preprint arXiv:2310.09909}, 2023.

\bibitem{xu2023multimodal}
Peng Xu, Xiatian Zhu, and David~A Clifton.
\newblock Multimodal learning with transformers: A survey.
\newblock {\em IEEE Transactions on Pattern Analysis and Machine Intelligence}, 2023.

\bibitem{yang2023dawn}
Zhengyuan Yang, Linjie Li, Kevin Lin, Jianfeng Wang, Chung-Ching Lin, Zicheng Liu, and Lijuan Wang.
\newblock The dawn of lmms: Preliminary explorations with gpt-4v (ision).
\newblock {\em arXiv preprint arXiv:2309.17421}, 9, 2023.

\bibitem{ye2023qilin}
Qichen Ye, Junling Liu, Dading Chong, Peilin Zhou, Yining Hua, and Andrew Liu.
\newblock Qilin-med: Multi-stage knowledge injection advanced medical large language model.
\newblock {\em arXiv preprint arXiv:2310.09089}, 2023.

\bibitem{zhang2023chatgpt}
Jizhi Zhang, Keqin Bao, Yang Zhang, Wenjie Wang, Fuli Feng, and Xiangnan He.
\newblock Is chatgpt fair for recommendation? evaluating fairness in large language model recommendation.
\newblock {\em arXiv preprint arXiv:2305.07609}, 2023.

\bibitem{zhao2023survey}
Wayne~Xin Zhao, Kun Zhou, Junyi Li, Tianyi Tang, Xiaolei Wang, Yupeng Hou, Yingqian Min, Beichen Zhang, Junjie Zhang, Zican Dong, et~al.
\newblock A survey of large language models.
\newblock {\em arXiv preprint arXiv:2303.18223}, 2023.

\bibitem{zhou2023comprehensive}
Hongyu Zhou, Xin Zhou, Zhiwei Zeng, Lingzi Zhang, and Zhiqi Shen.
\newblock A comprehensive survey on multimodal recommender systems: Taxonomy, evaluation, and future directions.
\newblock {\em arXiv preprint arXiv:2302.04473}, 2023.

\end{thebibliography}

% \input{content/05-supplementary}

%\end{refsection}
\end{document}